\pdfoutput=1
\documentclass[a4paper,11pt]{article}

\usepackage{jheppub}
\usepackage{graphicx}
\usepackage{amssymb,amssymb}

\usepackage{atlasphysics}
\usepackage{subfigure}
\usepackage{mathrsfs}
\usepackage{slashed}
\usepackage{setspace}
\usepackage{amsmath}
\usepackage{multirow}

\usepackage{xspace}

\newlength{\figwidth}
\newcommand{\POWHEG}{{\sc Powheg}}

\newcommand{\CTTen}{{\sc CT10}}
\newcommand{\MSTW}{{\sc MSTW2008}}
\newcommand{\NNPDF}{{\sc NNPDF2.0}}

\newcommand{\PYTHIA}{{\sc Pythia}}
\newcommand{\Hathor}{{\sc Hathor}}
\newcommand{\Herwig}{{\sc Herwig}}
\newcommand{\Jimmy}{{\sc Jimmy}}
\newcommand{\ACERMC}{{\sc AcerMC}}
\newcommand{\antikT}{anti-$k_{t}$}

\newcommand{\OVThree}{\ensuremath{OV_3}}
\newcommand{\htt}{HEPTopTagger\xspace}
\newcommand{\ttt}{Top~Template~Tagger\xspace}
\newcommand{\ca}{Cambridge/Aachen\xspace}

\newcommand{\secref}[1]{Section~\ref{#1}}

\def\MCatNLO{{\sc MC@NLO}}
\def\GEANT{{\sc Geant4}}
\newcommand{\gKK}{\ensuremath{g_{KK}}}
\newcommand{\xbr}{\ensuremath{\sigma\times  BR}}
\newcommand{\mtt}{\ensuremath{m_{\ttbar}}\xspace}
\newcommand{\CamKt}{\ensuremath{\mathrm{C/A}}\xspace}
\newcommand{\msubratio}{\ensuremath{m_{23}/m_{123}}\xspace}
\newcommand{\atanmsubratio}{\ensuremath{\arctan(m_{13}/m_{12})}\xspace}
\newcommand{\mjet}{\ensuremath{m_j}}

\newcommand{\httA}{U}
\newcommand{\httB}{V}
\newcommand{\httC}{W}
\newcommand{\httD}{X}
\newcommand{\httE}{Y}
\newcommand{\httF}{Z}
\newcommand{\myFigwidth}{0.65\linewidth}

\usepackage{preprintcover}   
\PreprintCoverPaperTitle{Search for resonances decaying into top-quark pairs using fully hadronic decays in $pp$\ collisions with ATLAS at $\boldmath{\sqrt{s}}$ = 7\,TeV}   
\PreprintIdNumber{CERN-PH-EP-2012-291}   
\PreprintCoverAbstract{A search for resonances produced in 7~\TeV\ proton--proton collisions
and decaying into top-quark pairs is described. In this Letter events where the  top-quark decay produces two massive jets with large transverse momenta recorded with the ATLAS detector at the Large  Hadron Collider are considered.
Two techniques that rely on jet substructure are used to separate top-quark 
jets from those arising from light quarks and gluons.
In addition, each massive jet is required to have evidence of an associated bottom-quark decay.
The data are consistent with the Standard Model, and 
limits can be set on the production cross section times branching fraction of 
a \Zprime\ boson and a Kaluza--Klein gluon resonance.
These limits exclude, at the 95\%\ credibility level, \Zprime\ bosons with masses 0.70--1.00~\TeV\ 
as well as 1.28--1.32~\TeV\ and 
Kaluza--Klein gluons with masses 0.70--1.62~\TeV.
}   
\PreprintJournalName{JHEP}   

\usepackage{mciteplus}

\title{Search for resonances decaying into top-quark pairs using fully hadronic decays in $pp$\ collisions with ATLAS at $\boldmath{\sqrt{s}}$ = 7\,TeV}

\author{The ATLAS Collaboration}

\date{today}

\abstract{
A search for resonances produced in 7~\TeV\ proton--proton collisions
and decaying into top-quark pairs is described. In this Letter events where the  top-quark decay produces two massive jets with large transverse momenta recorded with the ATLAS detector at the Large  Hadron Collider are considered.
Two techniques that rely on jet substructure are used to separate top-quark 
jets from those arising from light quarks and gluons.
In addition, each massive jet is required to have evidence of an associated bottom-quark decay.
The data are consistent with the Standard Model, and 
limits can be set on the production cross section times branching fraction of 
a \Zprime\ boson and a Kaluza--Klein gluon resonance.
These limits exclude, at the 95\%\ credibility level, \Zprime\ bosons with masses 0.70--1.00~\TeV\ 
as well as 1.28--1.32~\TeV\ and 
Kaluza--Klein gluons with masses 0.70--1.62~\TeV.
}

\keywords{Massive resonances, top quarks, top tagging, searches, boosted, substructure}

\begin{document}
\maketitle
\flushbottom

\setcounter{page}{1}

\section{Introduction}

Many models of new phenomena beyond the Standard Model (SM) predict resonances in the \TeV\
mass range that decay primarily into top-antitop quark pairs\footnote{In the following ``top quark'' refers to both the top quark and its anti-particle.} (\ttbar). 
This Letter reports on a search for such phenomena in 
proton--proton ($pp$) collisions at the Large Hadron Collider (LHC)
where both top quarks are reconstructed in 
their fully hadronic final states
and have large transverse momentum (\pT). 
The decay products of each high-\pT\ top quark 
are collimated and merge into one jet with large invariant mass.

Previous searches mostly considered cases where in one or both
of the top-quark decays, the intermediate $W$ boson decays leptonically and hence the top-quark decays  result in one or two isolated leptons,
missing energy from the neutrinos, and jets in the final 
state~\cite{Aaltonen:2007dia,cdf_physrevlett_100,Abazov:2008ny,Aaltonen:2009tx,Aaltonen:2011ts,Abazov:2011gv,Aad:2012LJets,ATLAS-boosted-semilep-2fb}.
The requirements of a well-identified charged lepton isolated from nearby hadronic energy deposits and missing transverse 
energy reject a large fraction of background from  multijet production.
However, 
difficulties arise in these final states when the top-quark decay particles are collimated, since leptons from the top-quark decay are 
no longer isolated and thus background contributions with lepton candidates originating from hadronic jets
are more difficult to distinguish from the signal.

An alternative approach that is reported in this Letter is to consider final states with 
high-\pT\ top quarks that decay hadronically and where the decay products are collimated 
in the direction of the  top-quark.
Such searches require the top quarks to have \pT\ in excess of 200--300~\GeV\ and
require rejection of the large background of gluon jets, light-quark jets, as well as $c$- and $b$-jets.
The CMS Collaboration employed this technique 
in a recent study~\cite{cmsboosted}.

In the present analysis, two complementary algorithms are 
used to identify top-quark decays and reconstruct the top-quark momentum for 
data collected with the ATLAS detector at a centre-of-mass energy of 7~\TeV.
The first algorithm is the \htt method~\cite{Plehn-htt,Plehn-stop} that 
tests the substructure of a jet reconstructed with the \ca (\CamKt) algorithm~\cite{Dokshitzer:1997}
with a large distance parameter $R=1.5$ (``fat jets'') for its compatibility with a hadronic top-quark decay.
This method is effective in identifying top-quark jets with \pT\ $>200$~\GeV.
The second algorithm is the \ttt method~\cite{Almeida2010:TopTemplate,Almeida2011:Higgs}\ that uses a large set 
of possible patterns of energy 
deposits (templates) from hadronic top-quark
decays to identify the best match to the observed energy deposits. 
The quality of the match is used to reject light quark and gluon jets. 
The \ttt\ uses jets reconstructed with the \antikT\ algorithm~\cite{Cacciari:2008gp}\ with a smaller distance parameter of $R=1.0$\ and 
is optimised to 
identify top quarks with $\pT > 450$~\GeV.
The invariant mass distributions of the \ttbar\ pair candidates identified using each
algorithm are examined  for evidence of resonance structure.

Two specific models that predict resonances of masses $m$\ with narrow and broad decay 
widths $\Gamma$ are considered: leptophobic topcolour \Zprime\ bosons
with  $\Gamma/m=1.2\%$~\cite{Harris:1999ya} 
and Kaluza--Klein (KK) gluons from the bulk Randall--Sundrum model (RS)\footnote{The
  left-handed ($g_L$) and right-handed ($g_R$) couplings to quarks in
  this model are: $g_L = g_R = -0.2 g_S$ for light quarks
  including charm, where $g_S=\sqrt{4 \pi \alpha_s}$; $g_L = g_S$ and $g_R = -0.2 g_S$ for 
bottom quarks; and $g_L = g_S$ and $g_R = 4 g_S$ for the top quark.}\
with $\Gamma/m = 15.3$\%~\cite{Agashe:2006hk,Lillie:2007yh,Lillie:2007ve}. 
The theoretical cross sections for the
\Zprime\ boson model and the bulk Randall--Sundrum model (RS)
are calculated with the \PYTHIA\ v6.421 MC generator \cite{Pythia8} and 
the {\sc Madgraph v4.4.51}~\cite{Alwall:2007st}\ MC generator, respectively.
A k-factor of 1.3 is applied to the \Zprime\ boson cross sections to account for NLO effects~\cite{Gao:2010bb}. 
Recent results from the 
ATLAS Collaboration in the lepton plus jets channel~\cite{Aad:2012LJets,ATLAS-boosted-semilep-2fb}\ 
exclude \Zprime\ bosons (KK gluons) 
with masses  0.5--1.15~\TeV\ (0.5--1.5~\TeV) at 95\%\ credibility level (CL). 
The CMS Collaboration obtained similar results~\cite{cmsboosted,cmssemilep} excluding 
0.50--1.49~\TeV\ for narrow ($\Gamma/m=1.2\%$) \Zprime\ signals,
0.50--2.04~\TeV\ for broad  ($\Gamma/m=10\%$) \Zprime\ signals,
and
1.00--1.82~\TeV\ for KK gluon signals.

This Letter is organised as follows: \secref{s:detector} describes the ATLAS 
detector and \secref{sec:samples} summarises the data samples and Monte Carlo (MC) event generators 
used in the analysis. 
The event selection and the definition of the reconstructed objects
are given in \secref{sec:selection}. The \htt and \ttt algorithms are described in 
\secref{sec:HEPTopTagger}\ and \secref{sec:TopTemplateTagger}, respectively. 
Estimates of the background rates and systematic uncertainties are given
in \secref{sec:bg} and \secref{uncertainties}, respectively. 
In~\secref{results} the resulting \ttbar\ mass spectrum  and exclusion limits are presented.

\section{ATLAS detector}
\label{s:detector}

The ATLAS detector~\cite{ATL-2008-001} at the LHC~\cite{LHCmachine} covers nearly the entire
solid angle\footnote{ATLAS uses a right-handed coordinate system with its origin at the nominal interaction point (IP) in the centre of the detector and the $z$-axis along the beam pipe. 
The $x$-axis points from the IP to the centre of the LHC ring, and the $y$-axis points upward. 
Cylindrical coordinates $(r,\phi)$ are used in the transverse $(x,y)$\ plane, $\phi$ being the azimuthal angle around the beam pipe. 
The pseudorapidity is defined in terms of the polar angle $\theta$ as $\eta=-\ln\tan(\theta/2)$. Distances in $(\eta,\phi)$ space are given as 
$\Delta R = \sqrt{(\Delta \phi)^2+(\Delta \eta)^2}$.}\ around the $pp$ collision point. 
The inner tracking detector 
(ID) comprises a silicon pixel detector, a silicon microstrip detector, and a transition radiation tracker, providing tracking capability within ${|\eta|<2.5}$. 
The ID is surrounded by a thin superconducting solenoid providing a 2~T axial 
magnetic field and by liquid-argon (LAr) electromagnetic sampling calorimeters with high granularity.
An iron/scintillator tile calorimeter 
provides hadronic energy measurements in the central rapidity 
range ($|\eta|< 1.7$). 
The end-cap and forward regions, covering $1.37<|\eta|<4.9$, are instrumented with LAr calorimeters for 
both electromagnetic and hadronic energy measurements.
The calorimeter system is surrounded by a muon spectrometer 
incorporating three 
superconducting toroid magnet assemblies. 

A three-level trigger system is used to select the events
for subsequent analysis. 
The level-1 trigger is implemented in hardware and uses a subset of
the detector information to reduce the rate to  at most
75~kHz. This is followed by two software-based trigger levels
that together reduce the event rate to a maximum of 400~Hz. 

\section{Data and Monte Carlo samples}
\label{sec:samples}

The analysis is performed using $pp$ collision data collected in 2011 
corresponding to an integrated luminosity of $4.7 \pm 0.2$~$\text{fb}^{-1}$~\cite{lumi,ATL-2011-002}.
With the increasing instantaneous luminosity of the LHC, the average number of simultaneous $pp$ interactions per 
beam crossing (pile-up) at the beginning of a given fill of the LHC
increased from about 6 to 17 during the 2011 data-taking period. 
The 2011 data pile-up conditions are included in the Monte Carlo simulation.

The main background contributions to a resonant signal in the \ttbar\ channel consist of SM \ttbar\ production and multijet events from gluon and non-top-quark production.
Fully hadronic SM \ttbar\ production
is simulated using the \MCatNLO\ v4.01 generator \cite{FRI-0201,FRI-0301}\ 
with \CTTen\ parton distribution functions (PDFs)~\cite{ct10} and assuming a top-quark mass of 172.5~\GeV. 
Final-state parton showers are simulated and hadronised using the \Herwig\ v6.5~\cite{Corcella:2002jc} program in association with the \Jimmy\ underlying event model~\cite{Butterworth:1996zw}.  
A \ttbar\ production cross section of 167~pb is used, calculated at approximate 
next-to-next-to-leading order (NNLO) in QCD 
using the \Hathor\ v1.2 Monte Carlo program~\cite{Aliev:2010zk}. 
This prediction employs the \MSTW\ NNLO PDF sets~\cite{mstwpdf}. 

The other background contributions, dominated by 
multijet events arising from the production of light quarks and gluons, 
but also  including smaller background contributions such as $W\!$+jets production and any
remaining contributions from \ttbar\ events where one of the top quarks decays
semileptonically (lepton+jet events), are estimated from data in 
signal-depleted control regions.
These are referred to as the multijet background in the following. 
Cross-checks of these background estimates are performed using \PYTHIA~\cite{PYTHIA} MC dijet samples.

Simulated signal samples for the $pp \rightarrow \Zprime \rightarrow t \bar{t}$ process
are produced using the \PYTHIA\ v6.421 MC generator with \MSTW\ PDFs~\cite{mstwpdf}. 
KK gluon final states are generated with the {\sc Madgraph v4.4.51}~\cite{Alwall:2007st}\ MC generator
with {\sc CTEQ6L1} PDFs~\cite{cteq6l} and using the \PYTHIA\ MC to model the parton shower and hadronization. 
These are calculated with leading-order matrix elements. Possible interference effects between the
\ttbar\ resonances and the SM \ttbar\ continuum are not taken
into account. 

The generated events are passed through a full simulation of the 
ATLAS detector~\cite{ATL-2010-005} based on \GEANT~\cite{AGO-0301}\ and then
processed with the same reconstruction algorithms used for the $pp$\ collision data events.
  
\section{Event selection and physics object reconstruction}
\label{sec:selection}
\label{sec:objreco}

The events for this analysis are selected with triggers matched to efficiently
identify collisions that meet the subsequent selection requirements.
The trigger for the \htt\ selection
uses the logical OR of two triggers based on jets defined using the \antikT\ algorithm with a distance
parameter $R=0.4$.
The first one requires the transverse energy ($\ET$) of at least one jet to satisfy
$\ET >100$~\GeV\ and the scalar sum of all jets to satisfy $\sum \ET >350$~\GeV\
($>400$~\GeV\ for later data-taking periods).
The second trigger requires at least five
jets with $\ET > 30$~\GeV. 
The combined single-jet and $\sum \ET$\ trigger is 
useful as it does not rely on the precise
topology of the \ttbar\ decay, which may change due to the splitting and
merging of jets, but relies mainly on the total energy deposited in the
calorimeter. The high-jet-multiplicity trigger is used to increase the
efficiency at low \ttbar\ invariant mass ($m_{\ttbar}$) where the top-quark decay products are often
reconstructed individually at trigger level.
The trigger for the \ttt selection requires an event to have at least 
one \antikT\ jet with a distance parameter $R=1.0$\ and $\ET > 240$~\GeV.

The events for both tagger selections are required to have a primary  vertex 
with at least five tracks with $\pT > 0.4$~\GeV. In the case of multiple vertex candidates
the primary vertex is defined as the one with the largest $\sum \pt^2$ of the tracks associated with it.

The analysis uses various jet-finder algorithms and distance parameters to reconstruct top-quark candidates and
to suppress background. 
These jets are formed from topologically-related calorimeter energy deposits
(`topoclusters')~\cite{TopoClusters,jespaper2010} using 
the {\sc FastJet} software \cite{Cacciari200657,fastjet}. 
The topoclusters are calibrated using the local cluster weighting method (LCW~\cite{ATLAS-CONF-2010-053}). 

Events for the \htt selection are required to contain at least two fat jets with
$\pt>200\GeV$ and $|\eta|<2.5$.
Each of these fat jets is subjected to the \htt algorithm (explained in detail in the following section),
which either rejects the jet as being incompatible with a hadronic top-quark decay or
reconstructs a top-quark candidate four-momentum.
To ensure high reconstruction efficiency, only top-quark candidates with $\pt > 200\GeV$ are considered in the following.

Events for the \ttt selection are required to have at least two jets 
reconstructed with the \antikT\ algorithm with a distance parameter of $R=1.0$, 
with one jet with $\pt>500~\GeV$\ and $|\eta|<2.0$,
and a second jet with $\pt>450~\GeV$\ and $|\eta|<2.0$.

In both selections, the leading and next-to-leading jets are required to satisfy one of the top-quark
tagging algorithms.
The \ttbar\ invariant mass is constructed  from the four-momenta of these two top-quark
candidates.

To further suppress background events in which multiple light-quark
and/or gluon jets satisfy the kinematic requirements, a neural-network-based
$b$-tagging algorithm is used~\cite{AdvBtagMV1}. 
This algorithm uses information on the impact parameter, the secondary
vertex, and the decay topology as its input.

Candidate $b$-quark jets are defined using the 
\antikT\ algorithm with a distance parameter $R=0.4$, with 
each jet calibrated to the energy scale of hadronic jets \cite{jespaper2010}.
These $b$-jets must satisfy the requirements $\pT > 25$~\GeV{} and
$|\eta|<2.5$. 
In addition, more than $75\%$ of the 
transverse momentum of the tracks associated with the jet must be carried by tracks with $\pT > 0.5$~\GeV{} 
originating from the primary vertex. 
In the \htt (\ttt) selection, the $b$-quark candidates must lie within $\Delta R = 1.4$ ($1.0$) of a fat jet axis such that
each tagged top-quark jet is associated with a unique $b$-quark tagged jet.
The $b$-tagging efficiency for $b$-jets from decays of high-\pT\ top quarks ranges from 50\%\ to 70\%, 
decreasing with increasing jet \pT\
because of the increasing collimation of the charged particles in the jet. 
With the same algorithm, about 3.5\% (7\%) of light-quark and gluon
jets are mistagged as $b$-jets at $\pT=200$~\GeV\ ($\pT=1$~\TeV).

Additional data quality criteria are applied, rejecting events that contain \antikT\ $R=0.4$ jets
that are identified as likely resulting from instrumental failure or non-collision
background (e.g. cosmic rays, beam gas and beam halo)~\cite{jespaper2010}.

The selected event samples are made complementary to samples used in
searches for \ttbar\ resonances in the lepton+jet and dilepton
channels by rejecting events that contain at least one isolated
electron (with $\pT>25\:\GeV$) or muon candidate (with $\pT>20\:\GeV$)~\cite{ATLASCONFsemilep5fb}.

\section{The \htt algorithm}
\label{sec:HEPTopTagger}

The \htt method is designed to reconstruct hadronically decaying
top quarks that are sufficiently boosted for their decay products to lie inside
a single fat jet. 
The performance of the \htt has been studied extensively
using ATLAS $pp$ collision data and simulated events~\cite{HTT-perfNote}.

The \htt method operates on a fat jet that has been constructed using the \CamKt jet algorithm.
The same algorithm is employed to re-cluster the fat jet constituents into subjets.
Previous studies~\cite{Butterworth-substruct} have shown that, compared to 
the $k_t$ and SISCone~\cite{SISCONE} jet finders, the \CamKt algorithm provides the 
best signal efficiency and background rejection in the presence of 
underlying event activity for top-quark taggers like the \htt. 
In the following 
the term ``top-quark candidate'' refers to the object resulting from the \htt procedure.  

The main steps of the method are described in the following;
for a detailed description see Ref.~\cite{Plehn-stop}. 
In a first phase, the input fat jet is split into subjets by undoing the last \CamKt clustering steps. This procedure is repeated until all subjet masses are  below $50\GeV$. These subjets form the basis of the substructure analysis.
All combinations of three subjets (``triplets'' in the following) 
are tested for compatibility with a hadronic top-quark decay using the following procedure. 
First, contributions from the underlying event and pile-up are removed in a filtering step: 
The \CamKt algorithm is re-run on the topoclusters of the triplet subjets with a distance parameter equal to 
half of the smallest pair-wise distance 
between the triplet subjets (but at most $0.3$), and only the resulting five most energetic 
subjets are kept; the
remaining activity is discarded.  More than three subjets are
potentially retained to account for possible QCD radiation in order to
improve the reconstruction of the top-quark decay.

The constituents of those five subjets are then re-clustered
exclusively~\cite{exclusiveKt,fastjet} into three subjets again using the \CamKt\
algorithm.
The reconstructed energy of the subjets is calibrated to the energy of the
incoming hadron jet using a simulation of the calorimeter response to
particle jets~\cite{jespaper2010}.
The three sub-jets are then tested for compatibility with being
products of a $t\rightarrow Wb \rightarrow q^\prime\bar{q}b$ decay,
using invariant mass ratios.
If the mass ratio requirements are met, the top-quark candidate four-momentum
is obtained by summing the four-momenta of the subjets. 
The invariant mass $m_{t}$ of the top-quark candidate is required to lie
in the range from 140 to 210~\GeV, otherwise this triplet is discarded. 
If a top-quark candidate is found in more than one triplet, only the one with its mass closest to the 
measured top-quark mass~\cite{Lancaster:2011wr}\ of $172.3\GeV$ is used. 

\begin{figure}
  \centering
  \subfigure[\label{sub:mpileup}]{\includegraphics[width=\myFigwidth]{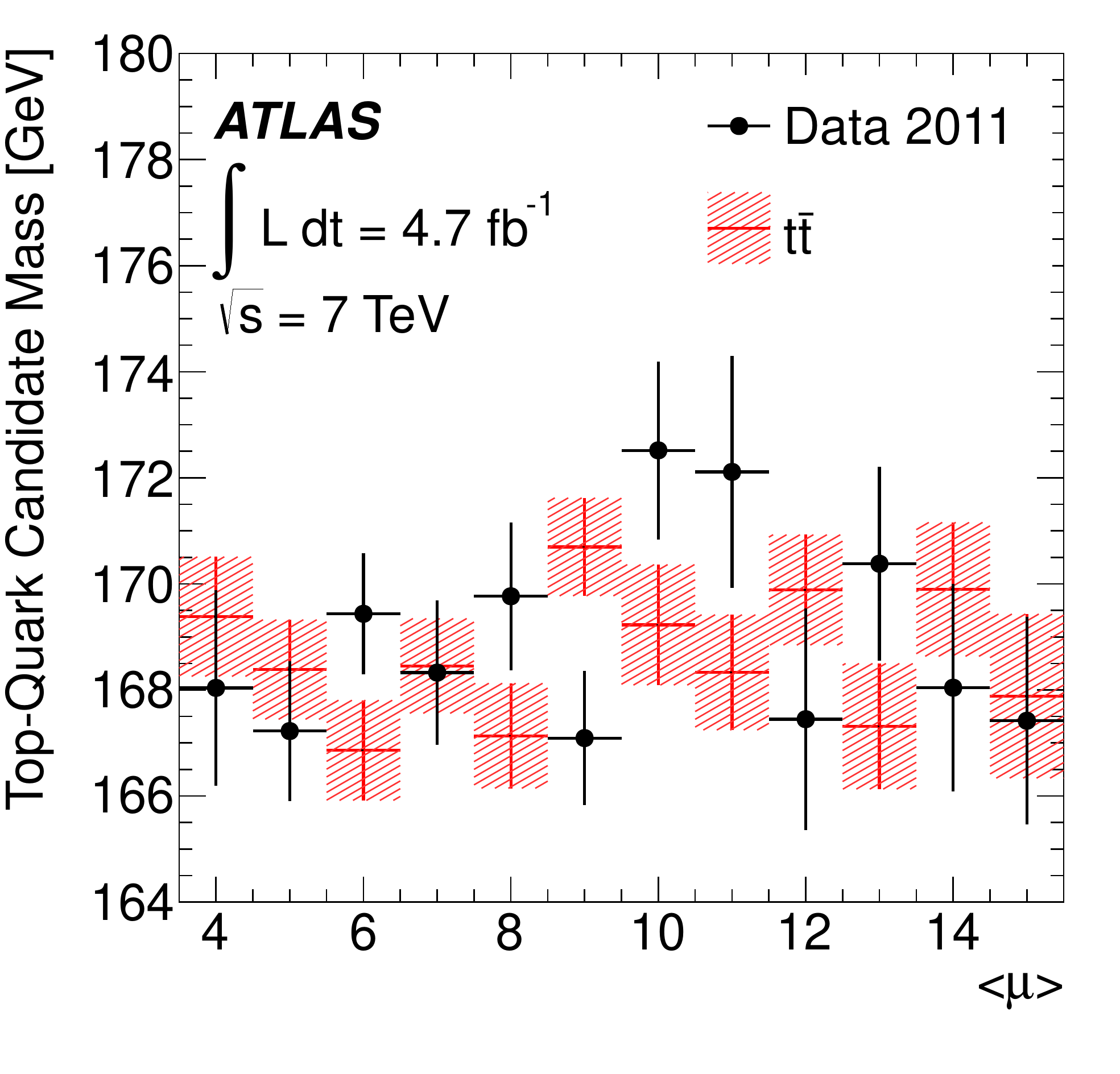}}
  \subfigure[\label{sub:mttpileup}]{\includegraphics[width=\myFigwidth]{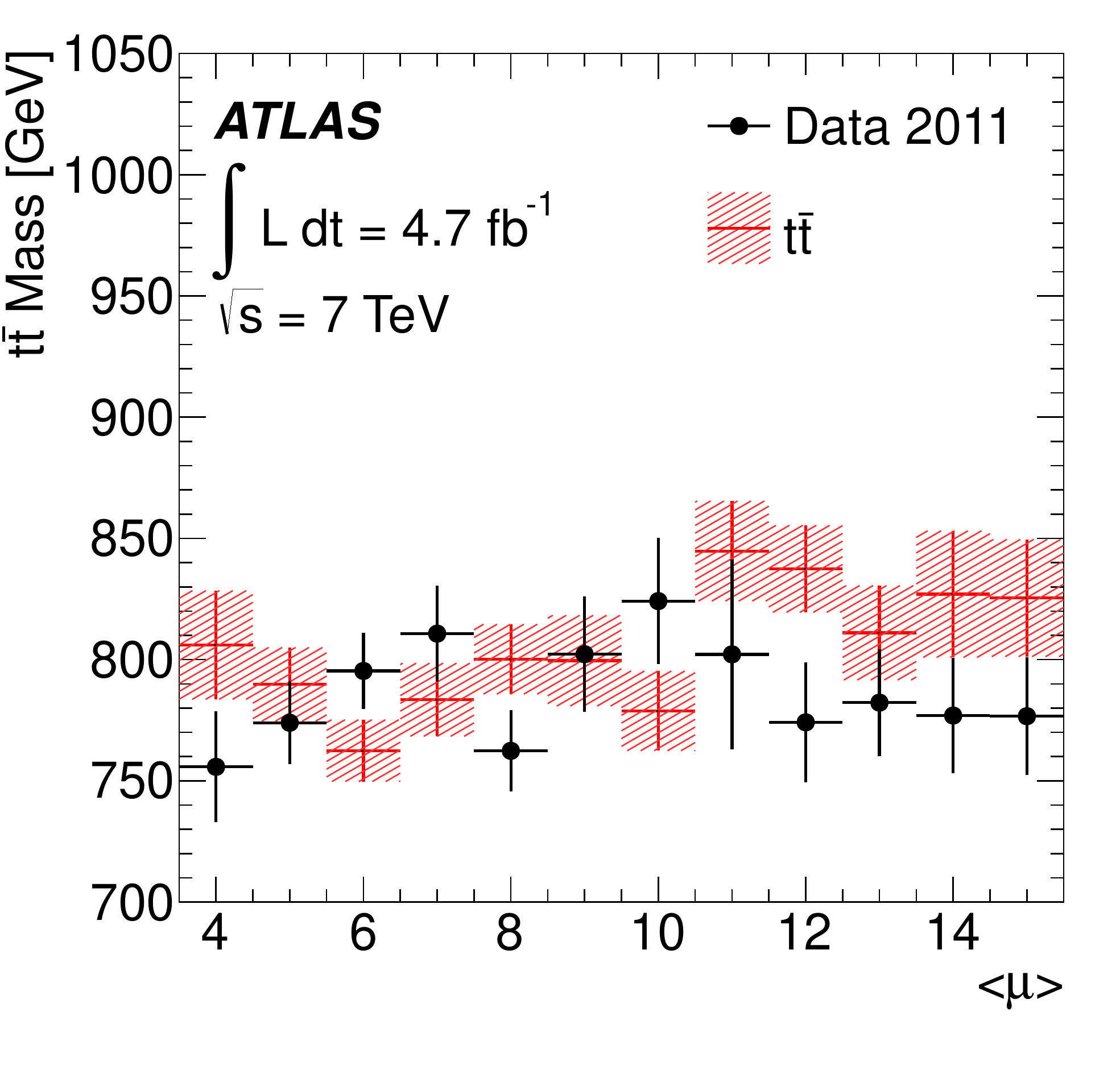}}
  \caption{Distributions of (a) mean \htt\ top-quark candidate mass and (b) mean reconstructed \ttbar\ mass
    as a function
    of the average number of interactions per bunch-crossing, $\langle \mu \rangle$, for data and simulated 
    \ttbar\ events with the full selection applied. Only statistical uncertainties are shown.}
  \label{fig:htt-pileup-mu}
\end{figure}

Distributions are shown in Fig.~\ref{fig:htt-pileup-mu}\ of
the mean reconstructed top-quark candidate mass~(a) 
and the reconstructed \ttbar\ mass averaged over the whole mass spectrum~(b)
as a function of the average number of interactions per bunch-crossing
for data and simulated \ttbar\ events.
The events are required to satisfy the \htt\ selection and to have two top-quark candidates.
No systematic shift of the mass with increased pile-up is observed within 
the statistical uncertainties.

\begin{figure}
  \centering
  \subfigure[\label{sub:zprimeshape}]{\includegraphics[width=\myFigwidth]{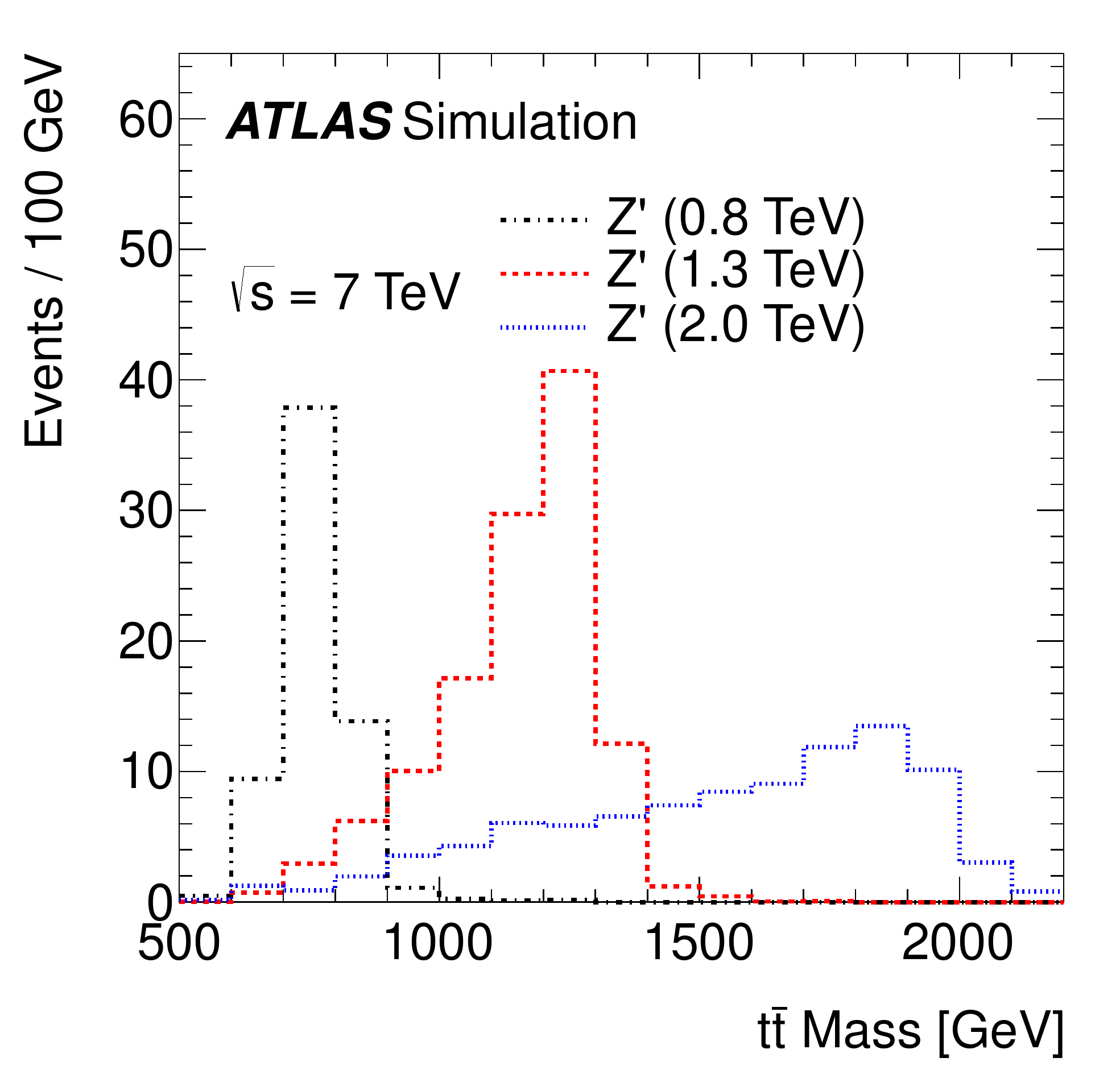}}
  \subfigure[\label{sub:kkgshape}]{\includegraphics[width=\myFigwidth]{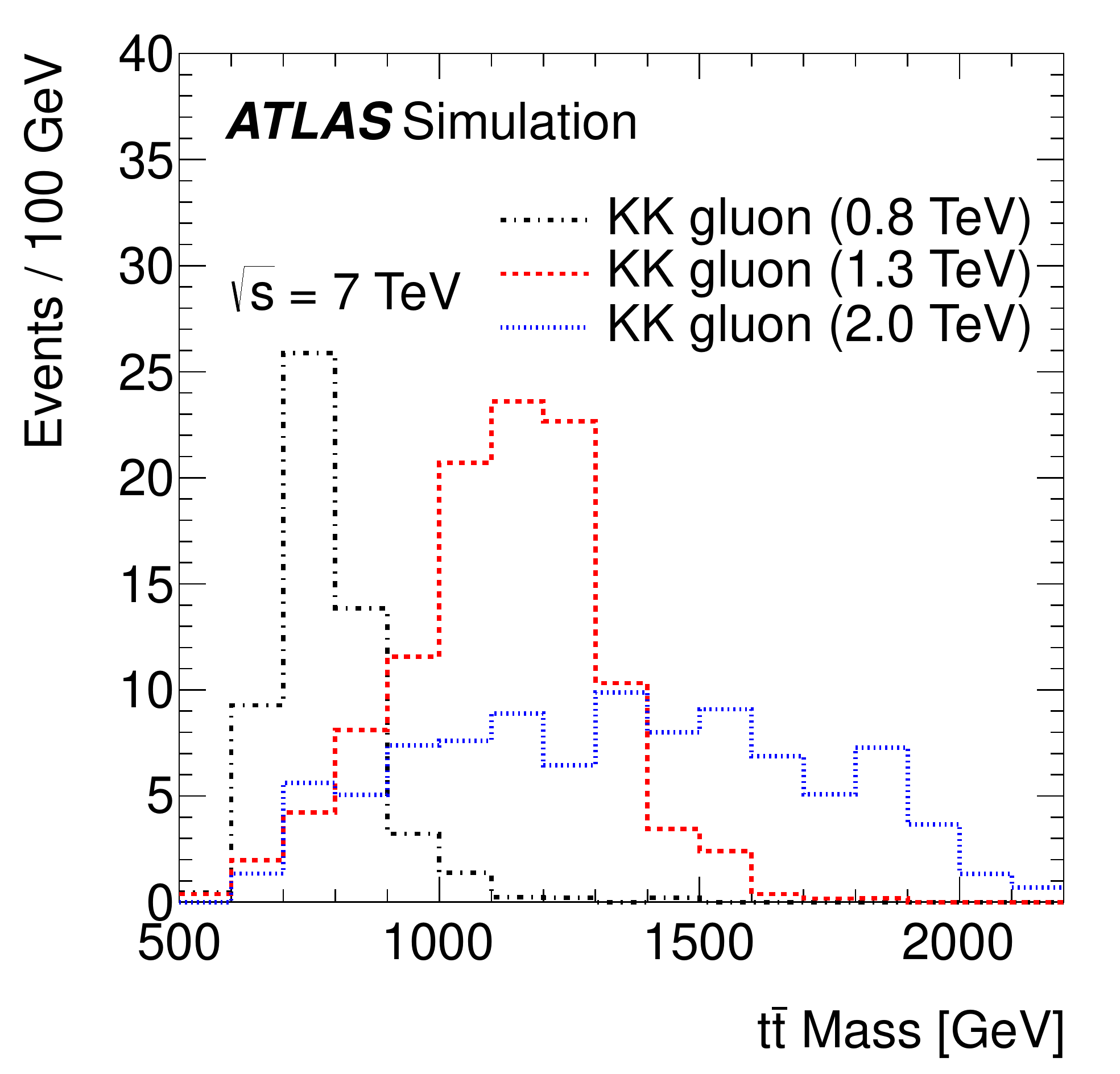}}

  \caption{Distributions of the reconstructed \ttbar\ mass predicted by MC simulations
    for (a) \Zprime\ boson and (b) KK gluon benchmark models with
    various mass values for the \htt analysis with the full selection applied.
    For each model, $\sigma(pp \rightarrow \Zprime/{\rm KK\:gluon}) \times BR(\Zprime/{\rm KK\:gluon} \rightarrow t\bar{t})$
    is fixed to $1\:\text{pb}$ and an integrated luminosity of $4.7\:\text{fb}^{-1}$ is assumed. }
  \label{fig:signal-shapes}
\end{figure}

The reconstructed \ttbar\ mass predicted by the MC simulations
for various \Zprime\ and KK~gluon masses is shown in
Fig.~\ref{fig:signal-shapes}. 

The total selection efficiency including both the \htt and $b$-tagging requirements is given in
Table~\ref{tab:effAcc} for various \Zprime\ boson and KK gluon masses, 
in events where the top quarks decay hadronically.  
The efficiency is dominated by the top-tagging and $b$-tagging efficiencies,
which vary as a function of the top- and bottom-quark momenta and are limited from above by

\begin{eqnarray}
\varepsilon_{b\text{-tag, max}}^{2} \cdot \varepsilon_{\text{top-tag, max}}^{2} \approx 10\%,
\end{eqnarray}

\noindent
where $\varepsilon_{b\text{-tag, max}}$\ is the maximum $b$-tagging efficiency of $80\%$ and 
$\varepsilon_{\text{top-tag, max}}$\ is the maximum top-tagging efficiency of $40\%$\ for
hadronically-decaying top quarks.
The efficiency drops for higher masses because of the decreasing $b$-tagging efficiency.

\begin{table}[!htb]
  \centering
  \caption{Total efficiency (in \%) for selecting \Zprime\ bosons and 
    KK~gluons ($g_{\text{KK}}$) that have
    decayed to \ttbar\ pairs.  
   These are the efficiencies determined by the MC calculations
   divided by the SM branching fraction of 46\% for both top quarks to decay 
   hadronically.
   All uncertainties are statistical only.}
  \vskip 0.2cm
  \begin{tabular}{|l|c|c|}
   \hline
    Model      & \multicolumn{2}{|c|}{Total Efficiency (\%)} \\
    \hline 
   & \htt & Template Tagger \\
   \hline

\Zprime\ ($0.5\, \TeV$) & $ 0.03  \pm 0.01 $ & -- \\
\Zprime\ ($0.8\, \TeV$) & $ 2.96  \pm 0.08$ & -- \\
\Zprime\ ($1.0\, \TeV$) & $ 4.76  \pm 0.09$ & $0.48 \pm 0.05$ \\
\Zprime\ ($1.3\, \TeV$) & $ 5.67  \pm 0.11$ & $6.37 \pm 0.13$ \\
\Zprime\ ($1.6\, \TeV$) & $ 5.40  \pm 0.10$ & $8.13 \pm 0.16$ \\
\Zprime\ ($2.0\, \TeV$) & $ 4.44  \pm 0.10$ & $6.26 \pm 0.13$ \\
& & \\
$g_{\text{KK}}$\ ($0.7\, \TeV$) & $ 1.70  \pm 0.13$ &  -- \\
$g_{\text{KK}}$\ ($1.0\, \TeV$) & $ 4.13  \pm 0.21$ & $0.74 \pm 0.10$ \\
$g_{\text{KK}}$\ ($1.3\, \TeV$) & $ 5.14  \pm 0.23$ & $5.02 \pm 0.25$ \\
$g_{\text{KK}}$\ ($1.6\, \TeV$) & $ 4.72  \pm 0.22$ & $6.43 \pm 0.26$ \\
$g_{\text{KK}}$\ ($2.0\, \TeV$) & $ 4.44  \pm 0.22$ & $5.22 \pm 0.21$ \\
    \hline
  \end{tabular}
 \label{tab:effAcc}
\end{table}

\section{The \ttt method}
\label{sec:TopTemplateTagger}

The \ttt method~\cite{Almeida2010:TopTemplate,Almeida2011:Higgs}\ is based on the concept
that an infrared-safe set of observables can be defined that quantify the overlap between
the observed energy flow inside a jet and the four-momenta of the partons 
arising from a top-quark decay.
An ``overlap function'' ranging from 0 to 1 is defined that quantifies the
agreement in energy flow between a given top-quark decay hypothesis (a template)  and an observed jet.  
One then cycles over a large set of templates chosen to cover uniformly the 3-body phase space for a 
top-quark decay at a given \pT\ and finds the template that maximises this 
overlap, denoted as \OVThree.
A requirement of $\OVThree>0.7$\ is made.

Sets (or ``libraries'') of approximately 300,000 templates are generated in steps of top-quark \pT\ of 100~\GeV\ starting from 450~\GeV\
by calculating the parton-level daughters for a top quark in its
rest frame and then boosting the daughters to the \pT\ of the given library. 
Studies of the top-quark jet tagging efficiency using MC data and of light quark/gluon jet rejection 
observed in the data were used
to determine the size of the \pT\ steps and the minimum number of templates for each library that maximise
the top-quark tagging efficiency while retaining high rejection against light quark/gluon jets.
For each jet candidate, the overlap function is defined as 

\begin{eqnarray}
\OVThree = 
\max_{\{\tau_n\}} \exp \Biggl[ 
- \sum_{i=1}^3 
 \frac{1}{2\sigma_i^2} \Bigl(E_i - \!\!\underset{<0.2}{\sum_{\Delta R(\mathrm{topo},i)}}\!\!\! E_{\mathrm{topo}} \Bigr)^{2} \Biggr], 
\end{eqnarray}

\noindent
where $\{\tau_n\}$\ is the set of templates defined for the given jet \pT, 
$E_i$\ are the parton energies of the top-quark decay daughters for the 
given template,
$E_{\mathrm{topo}}$\ is the energy of a topocluster, and
$\Delta R(\mathrm{topo},i)$\ is the $\eta-\phi$\ distance between the {$i^{th}$}\ parton and a
given topocluster.
The first sum is over the three
partons in the template and the second sum is over all topoclusters that are 
within $\Delta R(\mathrm{topo},i) = 0.2$\ and that have $\pT > 2$~\GeV.
The weighting variable is 

\begin{eqnarray}
\sigma_i = E_i/3.
\end{eqnarray}

\begin{figure}[h!]
   \centering 
         \includegraphics[width=\myFigwidth]{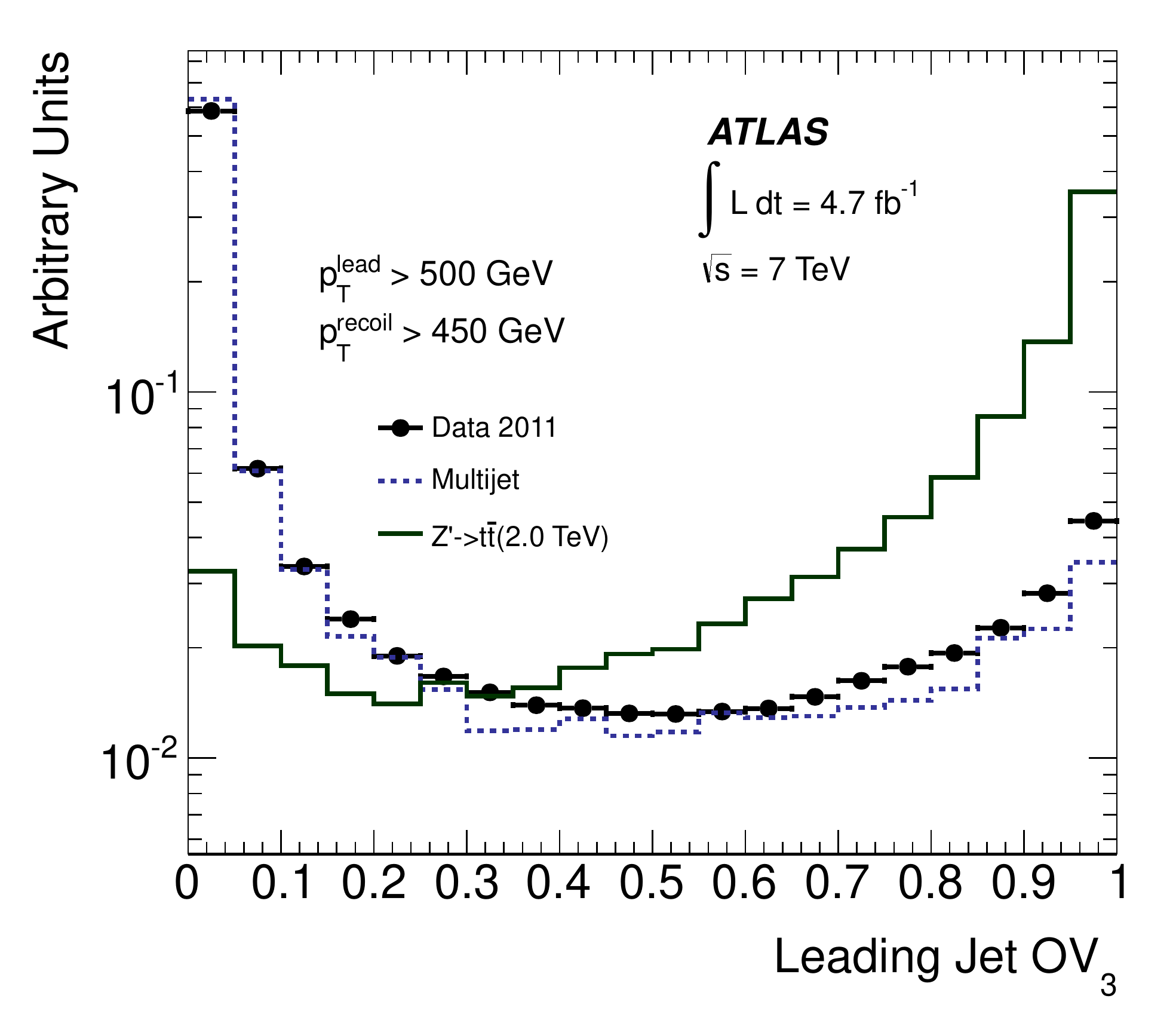}
  \caption{
The \OVThree\  distributions for the leading jets in the 2~TeV
$\Zprime \rightarrow \ttbar$\ MC sample, a multijet-dominated 2011 data sample, and
the multijet MC sample.
The data and multijet MC distributions are from the samples prior to
making any $b$-tagging or jet mass requirements on either jet, and so are
dominated by light quark/gluon jets.
}
\label{fig:OVThreeQCDttbar}
\end{figure}

The three tunable parameters in the \OVThree\ calculation --
the size of the cone used to match topoclusters with the parton,
the minimum \pT\ requirement on the topocluster, and
the weight $\sigma_i$\ -- have been determined from studies of the tagger's
performance judged by tagging efficiency and background rejection.
The overall performance is insensitive to the specific parameter values chosen.
The \OVThree\ distributions for a \Zprime\ MC sample, a multijet-dominated 2011 data sample, 
and the multijet MC sample are shown in Fig.~\ref{fig:OVThreeQCDttbar}, illustrating the separation of top-quark jets from the 
light quark/gluon jets in the large \OVThree\ region.

\begin{figure}
  \centering
  \subfigure[]{
    \includegraphics[width=\myFigwidth]{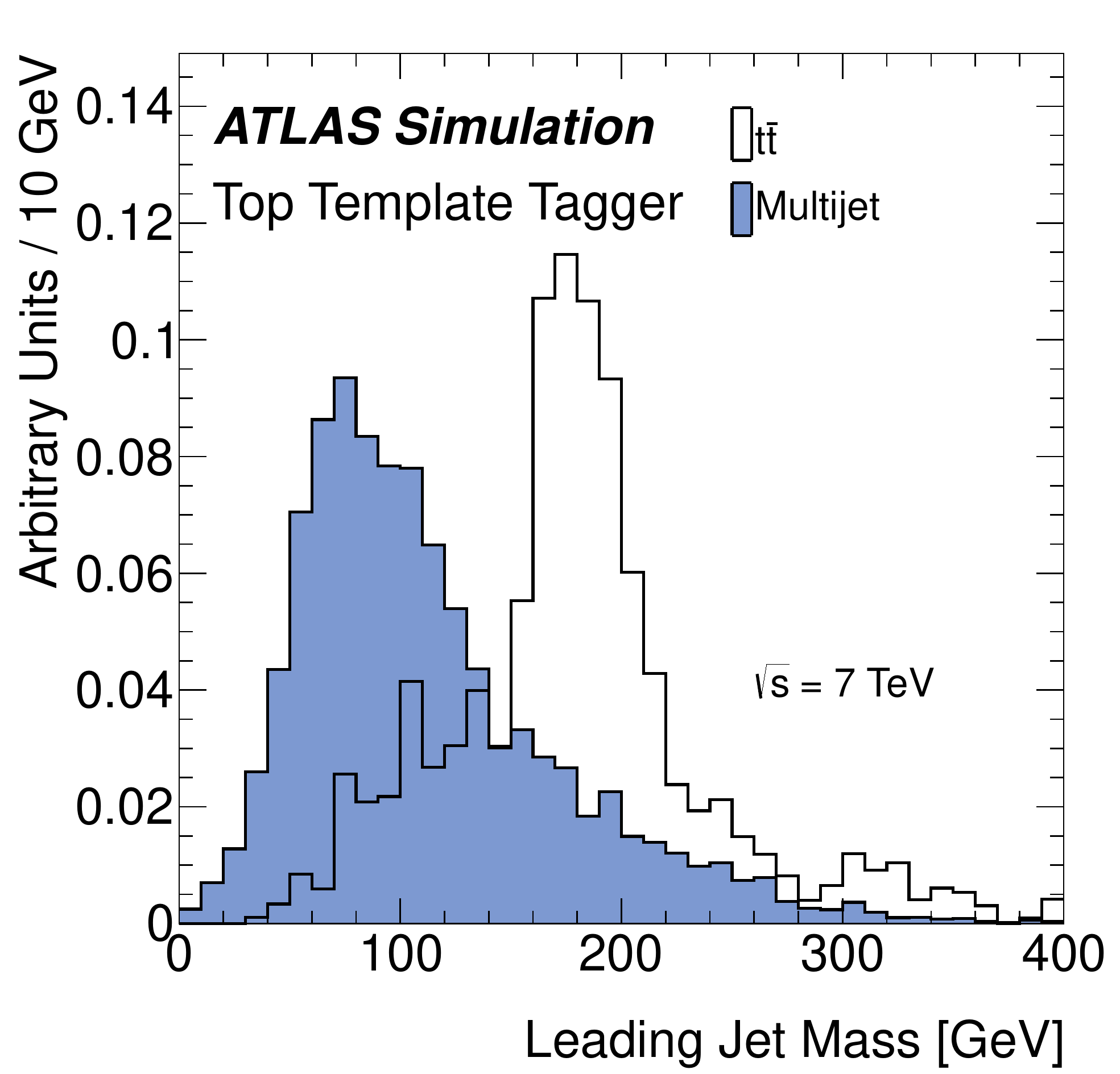}
}
  \subfigure[]{
    \includegraphics[width=\myFigwidth]{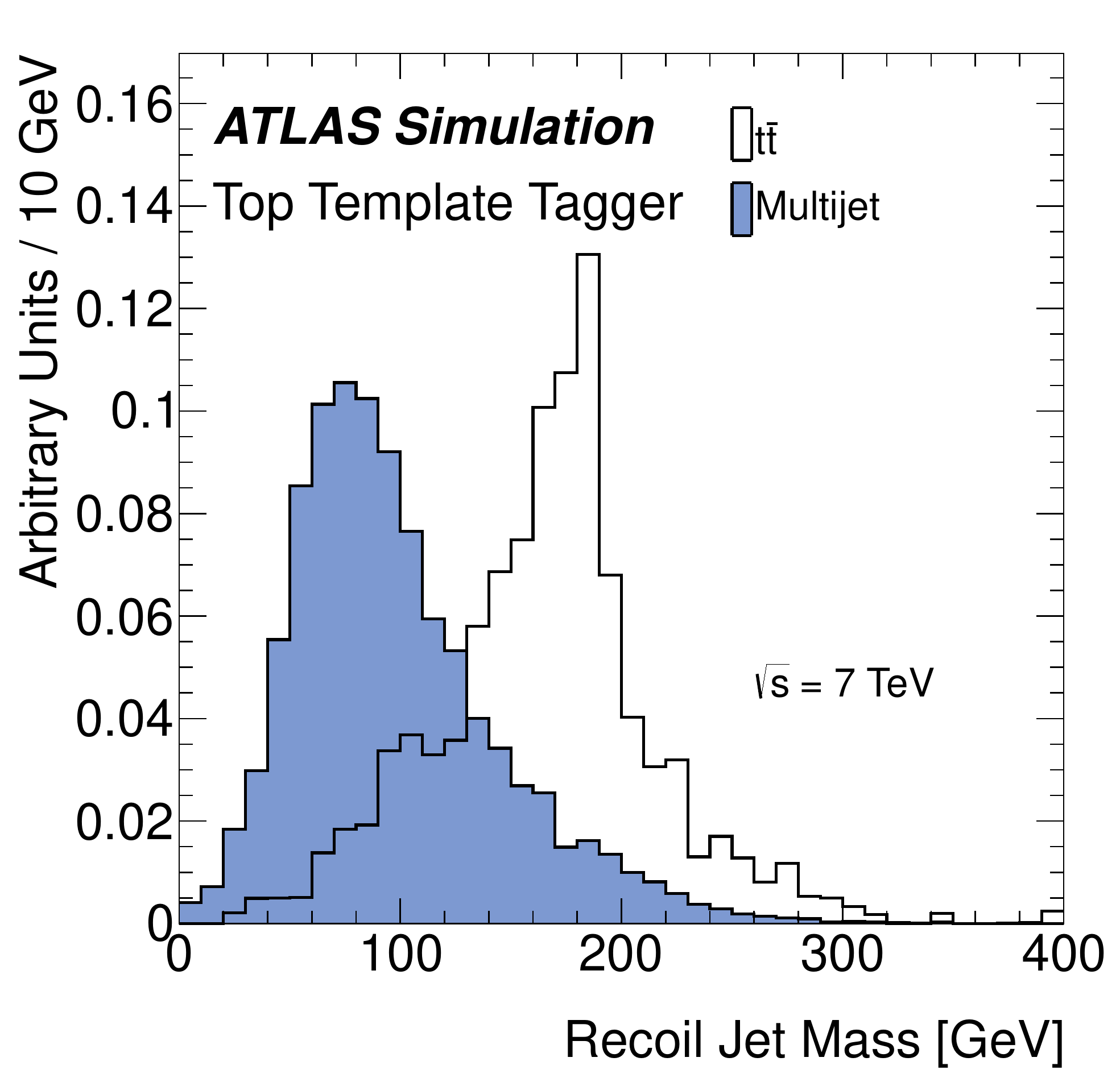}
}
  \caption{Pile-up-corrected jet mass distribution
in the multijet and \ttbar\ MC samples for (a) the leading and (b)
recoil jets.
In both cases, the jet mass requirement has been applied on the opposing jet in the event.
The distributions are independently normalised to unit area.
}
  \label{fig:jetmass-QCD-ttbar}
\end{figure}

The jet mass, \mjet, defined as the invariant mass of the topoclusters added
together as massless four-momenta~\cite{Blazey:2000qt}, 
has been shown to be an effective discriminant between top-quark jets and light quark/gluon
jets, even in the presence of multiple $pp$\ interactions~\cite{ATLAS:2012am,Aad:2012jf}.
A data-driven pile-up correction scheme 
for the jet mass is used, which measures the average mass shift 
experienced by jets using the flow of energy far from the jet as a function
of the number of multiple interactions in 
the event~\cite{Alon:2011xb,Aaltonen:2011pg}.
The discrimination of the pile-up-corrected jet mass between
light quark/gluon jets and top-quark jets is illustrated in Fig.~\ref{fig:jetmass-QCD-ttbar}\ 
for the leading and 
next-to-leading (or recoil) jet in the MC events that satisfy the \ttt selection.

The jet mass \mjet\ is required to be within $\pm50$~\GeV\ of the top-quark mass.

The jet mass and $\OVThree>0.7$\ requirements together have a rejection power
of $\sim10$\ for light quark/gluon jets that satisfy the kinematic requirements imposed on
the jets, based
on studies of samples dominated by light quark/gluon jets,
with an overall MC efficiency for selecting top-quark jets of $\sim75$\%.
Although \OVThree\ and \mjet\ are found to be correlated for a given jet,
the addition of the jet mass requirement increases the rejection
against light quark/gluon jets after an \OVThree\ requirement by a factor of two.
The combination of the \OVThree\ and \mjet\ requirements is 
therefore the core element of the \ttt.

\begin{figure}
  \centering
  \subfigure[]{
    \includegraphics[width=\myFigwidth]{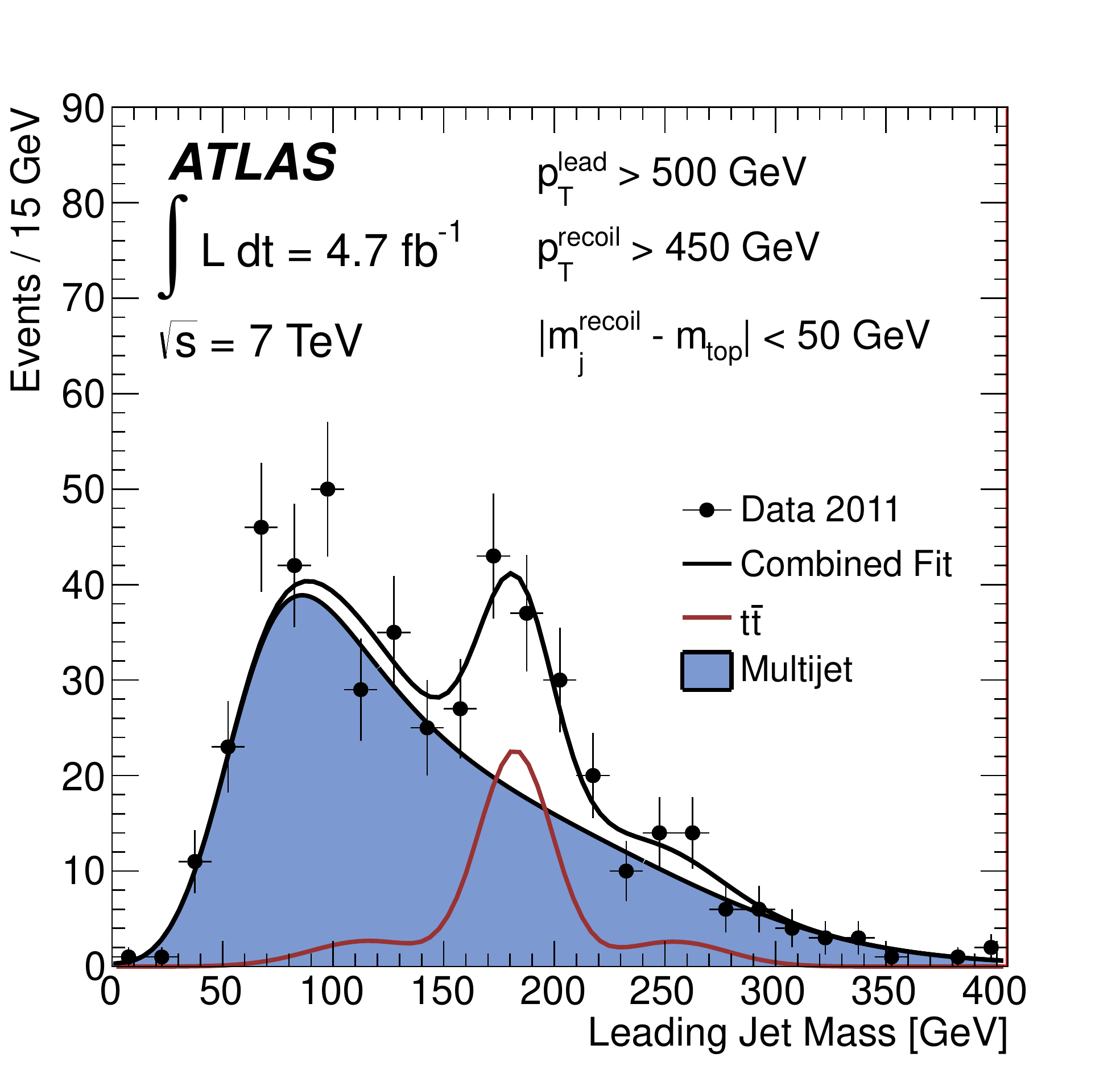}
}
  \subfigure[]{
    \includegraphics[width=\myFigwidth]{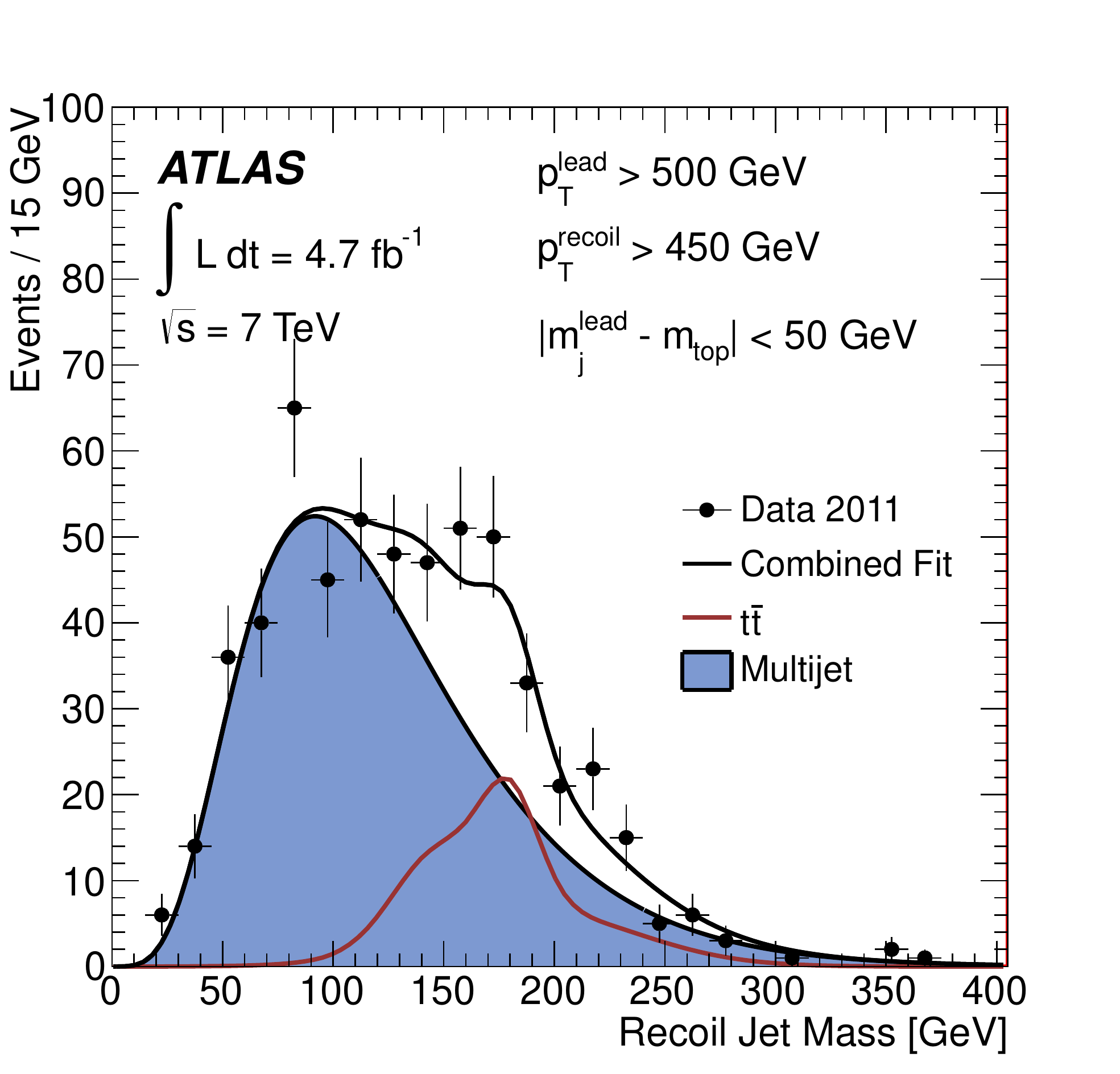}
}
  \caption{The jet mass distributions for the leading (a) and
for the recoil (b) jet when all other requirements have been made on the sample
except the mass and \OVThree\ requirements on the jet being considered.
The fits are described in the text.}
  \label{fig:tttjetmass}
\end{figure}

To verify that the tagger behaviour on top-quark jets is well modelled in 
the MC simulations,
an auxiliary analysis of the \ttt sample  is performed in which the \mjet\
and \OVThree\ requirements are relaxed on the leading jet.  
The resulting jet mass distribution, shown in Fig.~\ref{fig:tttjetmass}(a),
illustrates a clear peak from top-quark jets on top of a large background
from light quark/gluon jets.
The number of top-quark jets in this sample is measured by performing a 
fit to the background and top-quark jet signal, where the background shape
is determined from those events where the $b$-tag requirement has been
removed from the recoil jet and the top-quark signal shape is obtained from
the SM \ttbar\ MC simulations.
A smooth parameterisation has been used to describe the two distributions in the fit.
The number of top-quark jets that survive the jet mass and \OVThree\ requirements
on the leading jet is determined by subtracting the background in the 
signal region.
This results in a measured efficiency of the jet mass and \OVThree\ requirement
on top-quark jets of $0.81\pm0.25$, which is in agreement with the
estimate from the MC simulations of $0.75\pm0.07$\ (both statistical and systematic 
sources of uncertainty are included).

A similar analysis can be performed, interchanging the role of the leading jet and 
the recoil jet in the event.
This results in the jet mass distribution shown in Fig.~\ref{fig:tttjetmass}(b),
and in a top-quark tagging efficiency for the recoil jet of $0.62\pm0.20$, 
to be compared with the MC prediction of $0.62\pm0.05$.  

The overall efficiency of the \ttt selection on various signal samples is 
summarized in Table~\ref{tab:effAcc}.

\section{Background estimates}
\label{sec:bg}
The background contributions for both tagging analyses are estimated using control regions defined
by loosening the selection requirements for top-quark candidates and
for associated $b$-tagged jets.

\subsection{Background determination for the \htt analysis}
Six classes of events are created for the \htt analysis, as
outlined in Table~\ref{tab:regions}. 
They depend on the number of top-quark candidates and $b$-tagged jets. 
Regions \httE\ and \httF\ contain the events with at least two $b$-tags, with region \httE\ (\httF) 
additionally containing events with one (two or more) top-quark candidate(s). Region \httF\ constitutes the signal region. 

\begin{table}[hhh]
\caption{The classes of events used to calculate the data-driven
prediction for multijet background  events in the \htt analysis.
The numbers in parentheses are the estimated \ttbar\ purities in each region, given by the expected number of events arising 
from SM \ttbar{} production divided by the number of observed events in that region.}
\begin{center}
\begin{tabular}{|c|c|c|}

\hline
             & $1$ top-tag & $\geq 2$ top-tags \\
 \hline
 no $b$-tag    &   \httA (0.3\%)          &    \httB (2.4\%)    \\
 \hline
 1 $b$-tag     &  \httC (3.2\%)        &   \httD (24.3\%)     \\
 \hline
 $\geq 2$\ $b$-tags  &  \httE (22.5\%) &    \httF (80.9\%)    \\
 \hline
\end{tabular}
\end{center}
\label{tab:regions}
\end{table}

\begin{figure}[ht!]
  \centering
  \includegraphics[width=\myFigwidth]{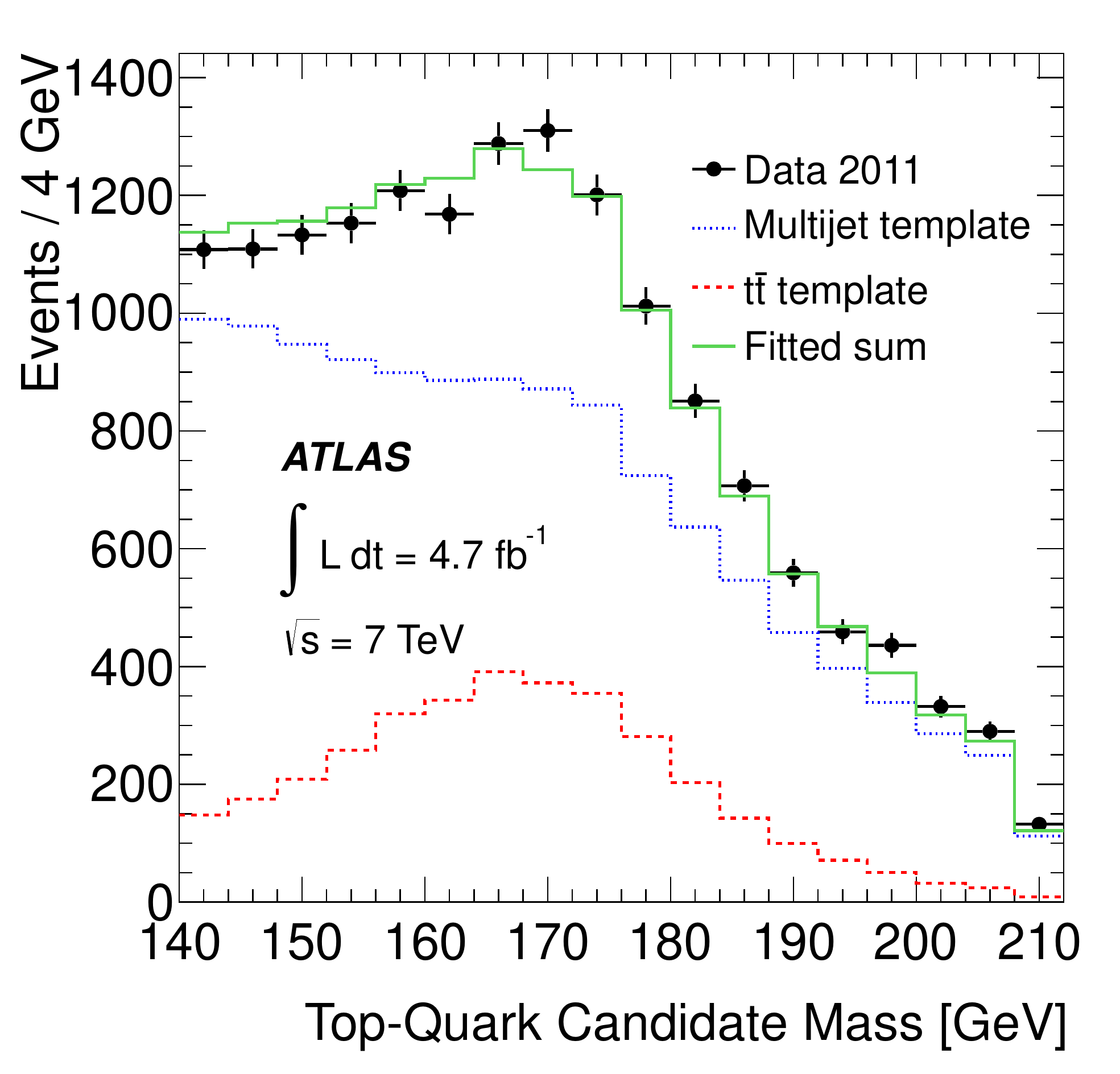}
  \caption{The distribution of the \htt top-quark jet candidate mass in the sideband region \httE\ for data, the templates 
for multijet background  and SM \ttbar{} production and the fitted sum.}
  \label{fig:htt-ttbar-template}
\end{figure}

The contribution of SM \ttbar\ production to each region is estimated from simulation and validated 
with data in region \httE\ as follows:
the top-quark candidate mass distribution in data, shown in Fig.~\ref{fig:htt-ttbar-template}, is 
fitted  with the sum of a \ttbar\ template
and a multijet background template,  
to extract the \ttbar\ background fraction, 
 exploiting the different shapes. 
The \ttbar\ template is taken from simulation. 
The multijet background template is defined as the data distribution in region \httC\ after subtracting the 
small contribution expected from SM \ttbar\ production in that region. 

The result is shown in
Fig.~\ref{fig:htt-ttbar-template}.  
The selection of the top-quark candidate closest in mass to the top-quark mass when multiple top-quark candidates
are reconstructed causes a small bias in the multijet background distribution, as seen in the figure.
The ratio of the fitted
\ttbar{} event yield to the predicted yield is
$1.01\pm0.09$, where the uncertainty is statistical.
This ratio is used to correct the normalisation of the SM \ttbar\ contribution in the 
determination of the multijet background in the signal region.
The resulting SM \ttbar\ yield in signal region \httF\ is estimated to be
$770 ^{+ 220 } _{ -180 }~\text{(stat.$\oplus$syst.)}$\ events. 

The multijet background is estimated by exploiting the fact that the number of $b$-tags and the 
number of top-quark tags are 
uncorrelated for this background.\footnote{The \htt does not use $b$-tagging information internally
and hence the probability for a multijet background event to fake a top-quark signal
is independent of the probability for it to fake a $b$-quark signal. This is verified using dijet MC samples.}
The shape of the multijet background for a given variable (e.g. $m_{\ttbar}$) is estimated from
the weighted average of the distribution of that variable in regions \httB\ and \httD, normalised by the yields in regions \httA\ and \httC\, respectively,
and scaled by the event count in region \httE:

\begin{eqnarray}\label{eq:n_QCD}
  \frac{\text{d}n_{\httF}}{\text{d}m_{\ttbar}} = \left(\frac{1}{n_{\httA}} \times \frac{\text{d}n_{\httB}}{\text{d}m_{\ttbar}}+\frac{1}{n_{\httC}}\times \frac{\text{d}n_{\httD}}{\text{d}m_{\ttbar}}\right) \times \frac{n_\httE}{2},
\end{eqnarray}

\noindent
in which $n_i$ is the number of events in region $i$ after subtracting  
the expected SM \ttbar\ background normalised to the observed \ttbar\ yield. 
Hence the \ttbar{} and multijet background contributions are anti-correlated.
The resulting estimate for the multijet background in the signal region is 
$130 \pm 70~\text{(stat.$\oplus$syst.)}$\ events. 

\begin{figure}
  \centering
  \subfigure[\label{sub:fatmass}]{\includegraphics[width=\myFigwidth]{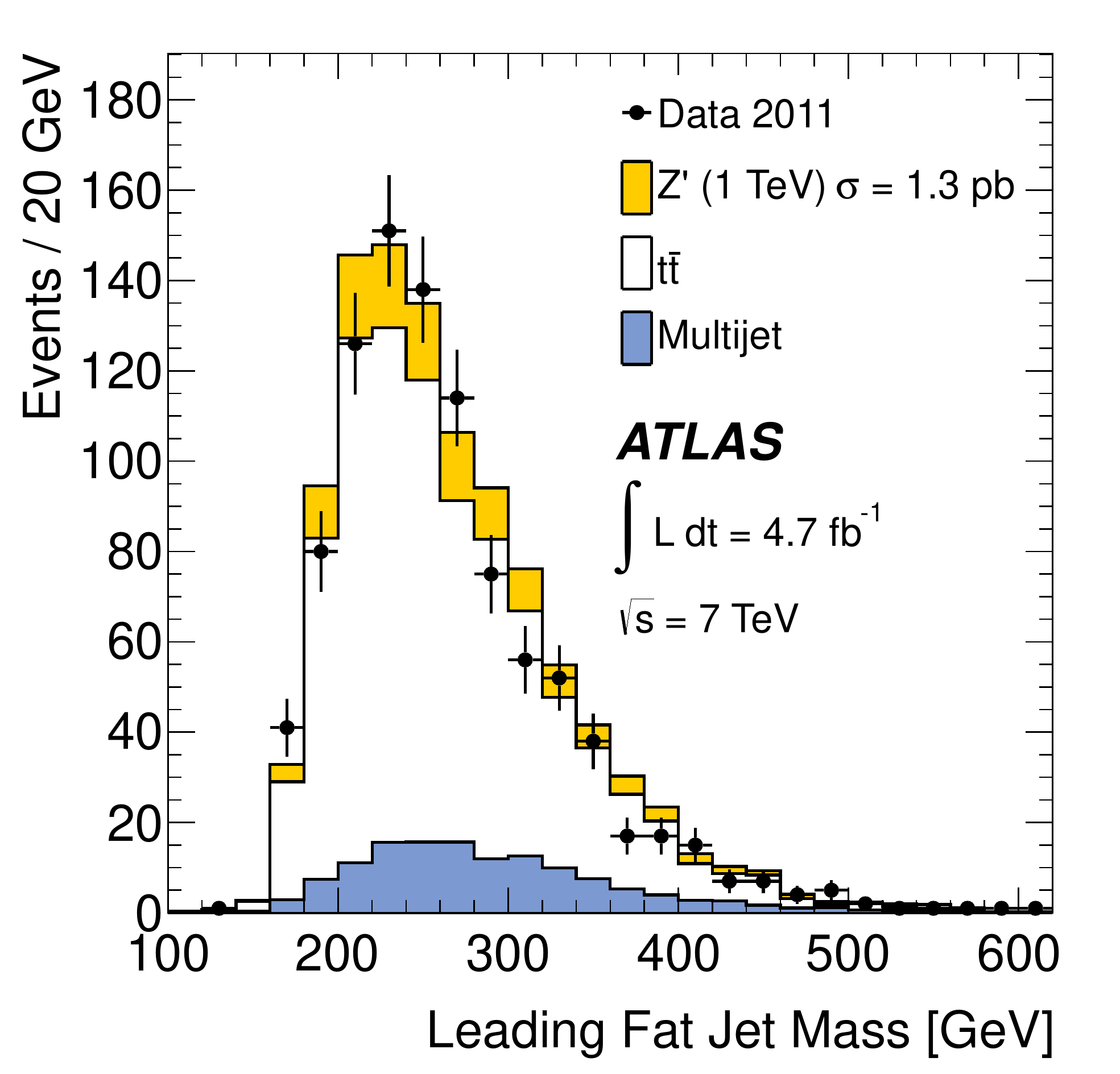}}\\
  \subfigure[\label{sub:topmass}]{\includegraphics[width=\myFigwidth]{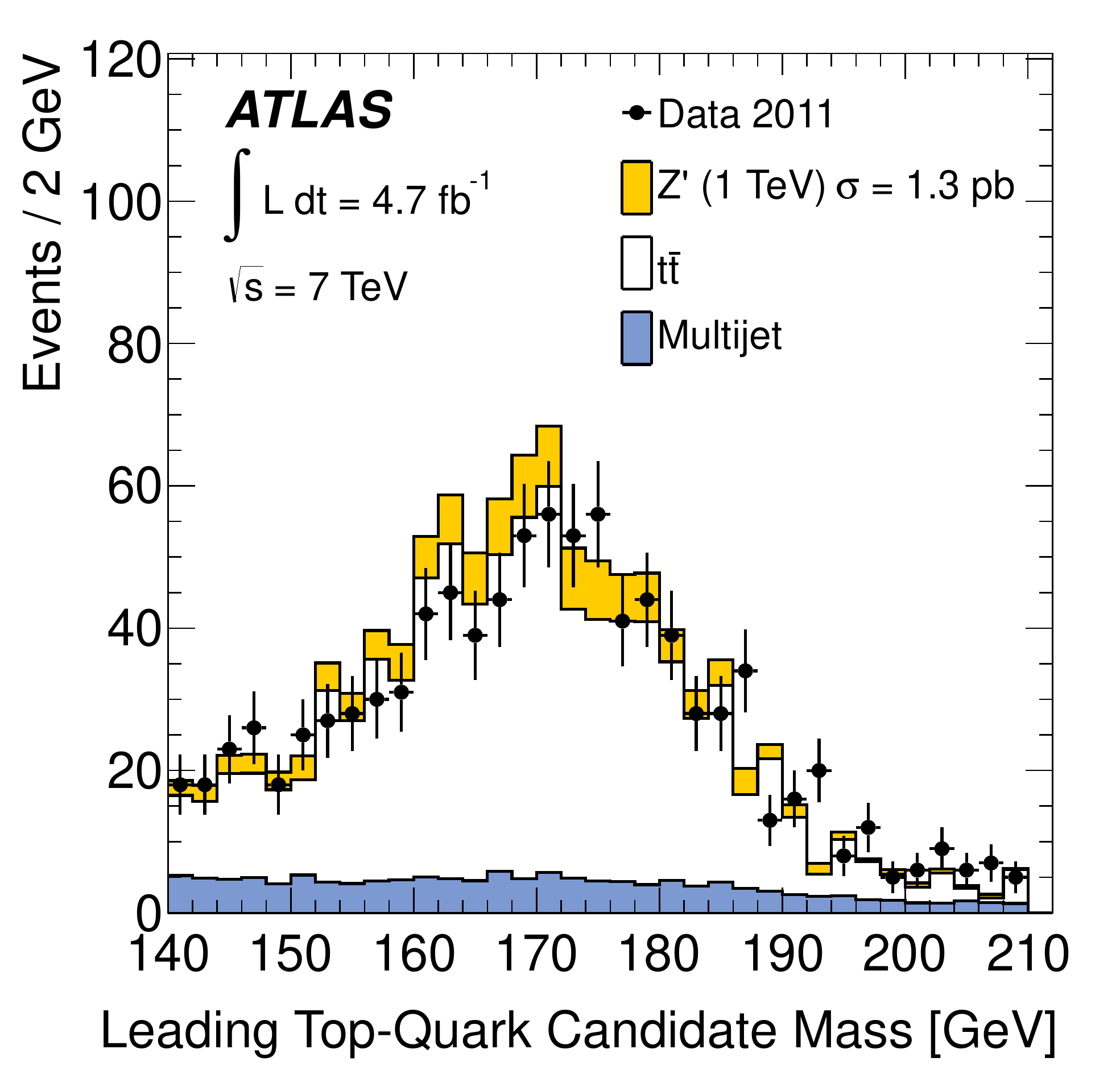}}\\
  \caption{Signal region distributions of (a) the mass of the leading \pt{} fat jet and (b) the mass of the leading \pt{} top-quark candidate. 
    Also shown are the prediction for SM \ttbar{} production, the multijet background contribution as 
estimated from data, 
    and a hypothetical \Zprime\ boson signal.}
  \label{fig:perfm}
\end{figure}

\begin{figure}
  \centering
  \subfigure[\label{sub:m23m123}]{\includegraphics[width=\myFigwidth]{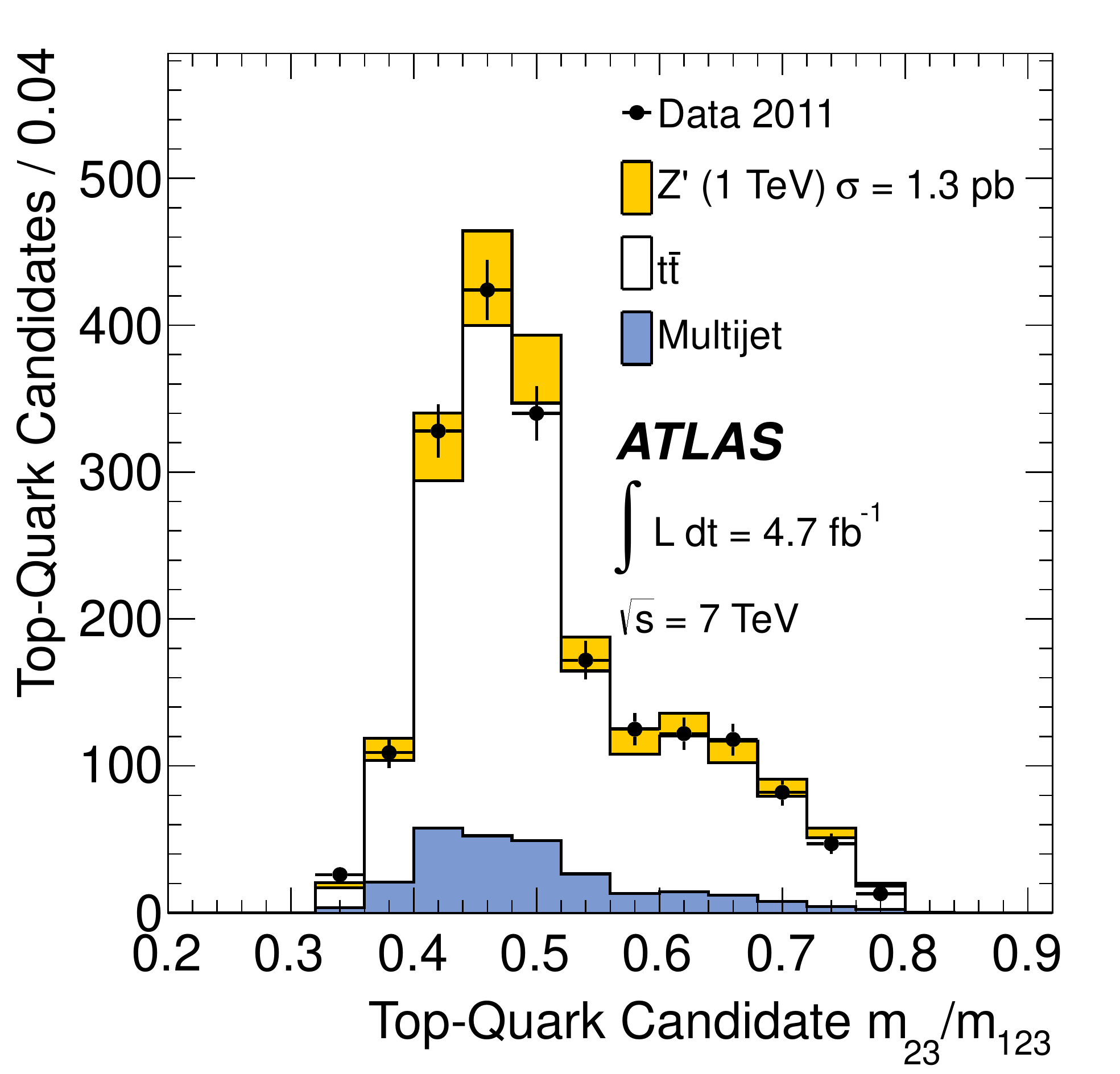}}\\
  \subfigure[\label{sub:atan1312}]{\includegraphics[width=\myFigwidth]{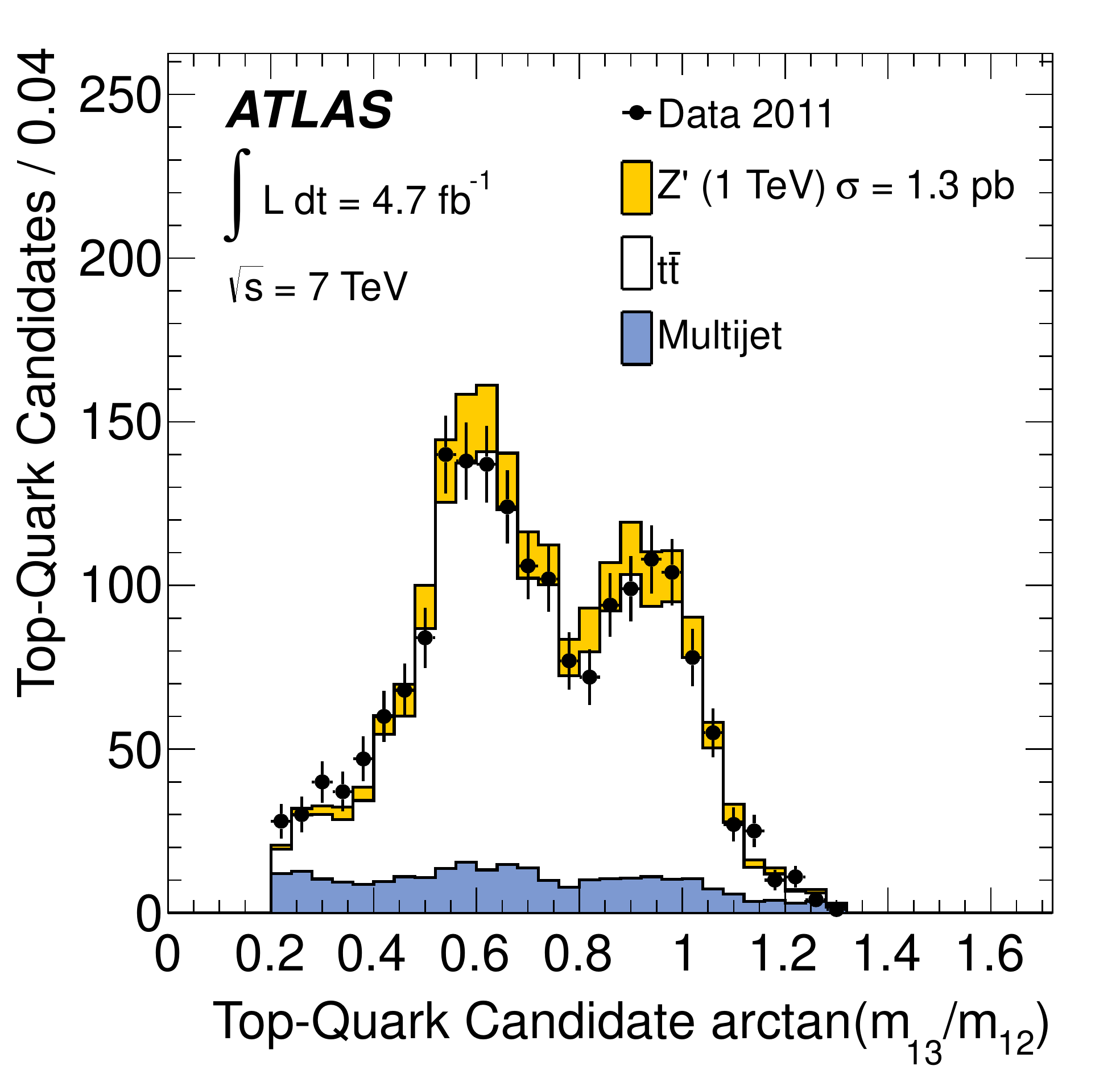}}\\
  \caption{Signal region distributions of the top-quark candidate substructure variables  \msubratio\ (a) and
    \atanmsubratio\ (b). Also shown are the prediction for SM \ttbar{} production, the multijet background
 contribution 
    as estimated from data, and a hypothetical \Zprime\ boson signal.}
  \label{fig:perfsub}
\end{figure}

To check that the multijet and SM \ttbar\ background predictions are consistent with the data and to 
illustrate that the \htt\ identifies top-quark jets effectively,
Figs.~\ref{fig:perfm}\ and~\ref{fig:perfsub}\ show comparisons of  predicted and 
observed distributions in the signal region: of the fat-jet mass (Fig.~\ref{sub:fatmass}), 
the top-quark candidate mass (Fig.~\ref{sub:topmass}), 
and  the  substructure variables 
\msubratio (Fig.~\ref{sub:m23m123}) and \atanmsubratio (Fig.~\ref{sub:atan1312}). 
In these ratios $m_{123}$ is the invariant mass of all
 three subjets and $m_{ij}$ is the invariant mass of subjets $i$ and $j$, where
the subjets have been sorted by  \pt\ in descending order.
The data are consistent with the sum of the multijet and SM \ttbar\ background predictions for all distributions. 

\subsection{Background determination in the \ttt analysis}

The multijet background for the \ttt analysis is estimated in a manner similar to the 
\htt analysis.  
Various control regions are used in order to reduce 
biases resulting from the observed correlations in \ttt tagging efficiencies between the recoil
and leading jet.

\begin{figure}
  \centering
  \includegraphics[width=\myFigwidth]{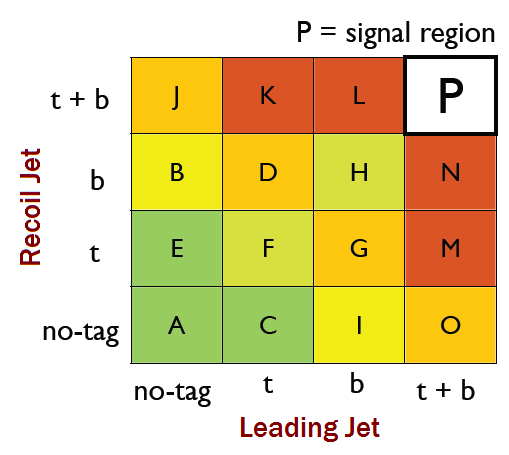}
  \caption{The 16 subsamples into which the \ttt data are divided, based on whether the leading
    and recoil jets have a $b$-quark tag, and on  whether they satisfy the Top Template tag
    requirements of $\OVThree > 0.7$.
    The jet mass requirement of $|\mjet - m_{t}|< 50$~\GeV\ is applied to both
    jets for all subsamples.
    The colour coding (in the online version) reflects the anticipated level of expected signal from both
    SM \ttbar\ production and possible production of \ttbar\ states
    through resonant production: $<0.25$\%\ (light green: A,C,E), 
    $0.25--10$\%\ (shades of yellow: B, D, F--J, O), and
    $> 10$\%\ (red: K--N).
  }
  \label{fig:Background-Signal-Matrix}
\end{figure}

The sample of events in the \ttt analysis prior to requiring either top-quark tags or $b$-quark tags
is divided into 16
discrete and non-overlapping subsamples, as shown in Fig.~\ref{fig:Background-Signal-Matrix}.
The jet mass requirement has been applied to both the leading and recoil jets in all subsamples.
An expected correlation in the masses of the leading and recoil jets~\cite{Blum:2011} leads 
to a non-negligible correlation in the top-quark tagging efficiency for the two jets in
dijet events. 
On the other hand, the $b$-quark tagging efficiency of the two jets is uncorrelated.
Jets produced from $b\bar{b}$\ pairs would create a small
correlation, but their overall rate is expected to be negligible in the samples used
below to calculate the multijet background.

The rate of multijet background events in the signal region (subsample P) is calculated with an iterative 
method that uses the lack of correlation in $b$-tagging efficiencies between the leading and
recoil jets. 
In its simple form, a two-dimensional-sideband counting 
technique for background estimation requires events to be selected using pairs of uncorrelated 
variables. For example, in our subsample grid, the top-tagging state of the leading jet is not 
correlated to the $b$-tagging state of the recoil jet in multijet background events. 
Therefore, the ratio of background events in region D to region C should be the same as the ratio of 
background events in region B to region A. 
This relation can be used to predict the background 
rate in region D using the observed rates in the other three regions.
The predicted number of SM \ttbar\ events in each subsample (which is of 
order 1\%\ or less for each region used in the background
calculation) is subtracted before this calculation is performed.

A number of the subsamples (regions K, L, M, 
and N) can contain potential \ttbar\ 
contributions from beyond-the-SM processes and therefore cannot be used in this method.
Furthermore, the AJOP grid cannot be used to predict the background rate in region P, due to 
the correlation in the top-tagging rates for the leading and recoil jets. 
An iterative calculation is performed:
background rates in subsamples K and M are determined with subsamples not potentially 
contaminated with top-quark jets, and these predicted 
rates are then used in a subsequent step to predict the background rate in the \ttt signal region:

\begin{eqnarray}
K^\prime &=& N_J\times \frac{N_F}{N_E} \\
M^\prime &=& N_F\times \frac{N_O}{N_C} \\
P^\prime &=& K^\prime\times \frac{M^\prime}{N_F} = {{\frac{N_J \times N_O \times N_F}{N_E \times N_C} }}, 
\end{eqnarray}

\noindent
where the $N_X$\ in these equations are the observed number of events in subsample $X$\
and $K^\prime$, $M^\prime$, and $P^\prime$\ are the predicted multijet background contributions in 
the associated subsample.

\begin{table}[!htb]
  \centering
  \caption{Results of the different predictions for the multijet background rates in the \ttt signal region.  
    The table lists the calculation performed and the corresponding predicted number of multijet background events.
    The uncertainties shown are statistical.}
  
  \vspace{0.3cm}
  \begin{tabular}{|l|c|}
    \hline
    Subsamples & Predicted Events  \\
    \hline
    $(J \times F \times O)/(E \times C)$ & $51 \pm 3$ \\
    \hline
    $(J \times F \times H \times O)/(E \times D \times I)$ & $56 \pm 6$ \\
    \hline
    $(J \times F \times H \times O)/(B \times C \times G)$ & $54 \pm 6$ \\
    \hline
    $(J \times F \times I \times O)/(A \times C \times G)$ & $51 \pm 4$ \\
    \hline
    $(J \times F \times B \times O)/(A \times E \times D)$ & $52 \pm 4$ \\
    \hline
    ${\bf{Average}}$ & ${\bf{53 \pm 3}}$ \\
    \hline 
  \end{tabular}
  \label{tab:ttt-signal-predictions}
\end{table}

\begin{figure}
  \centering

  \includegraphics[width=\myFigwidth]{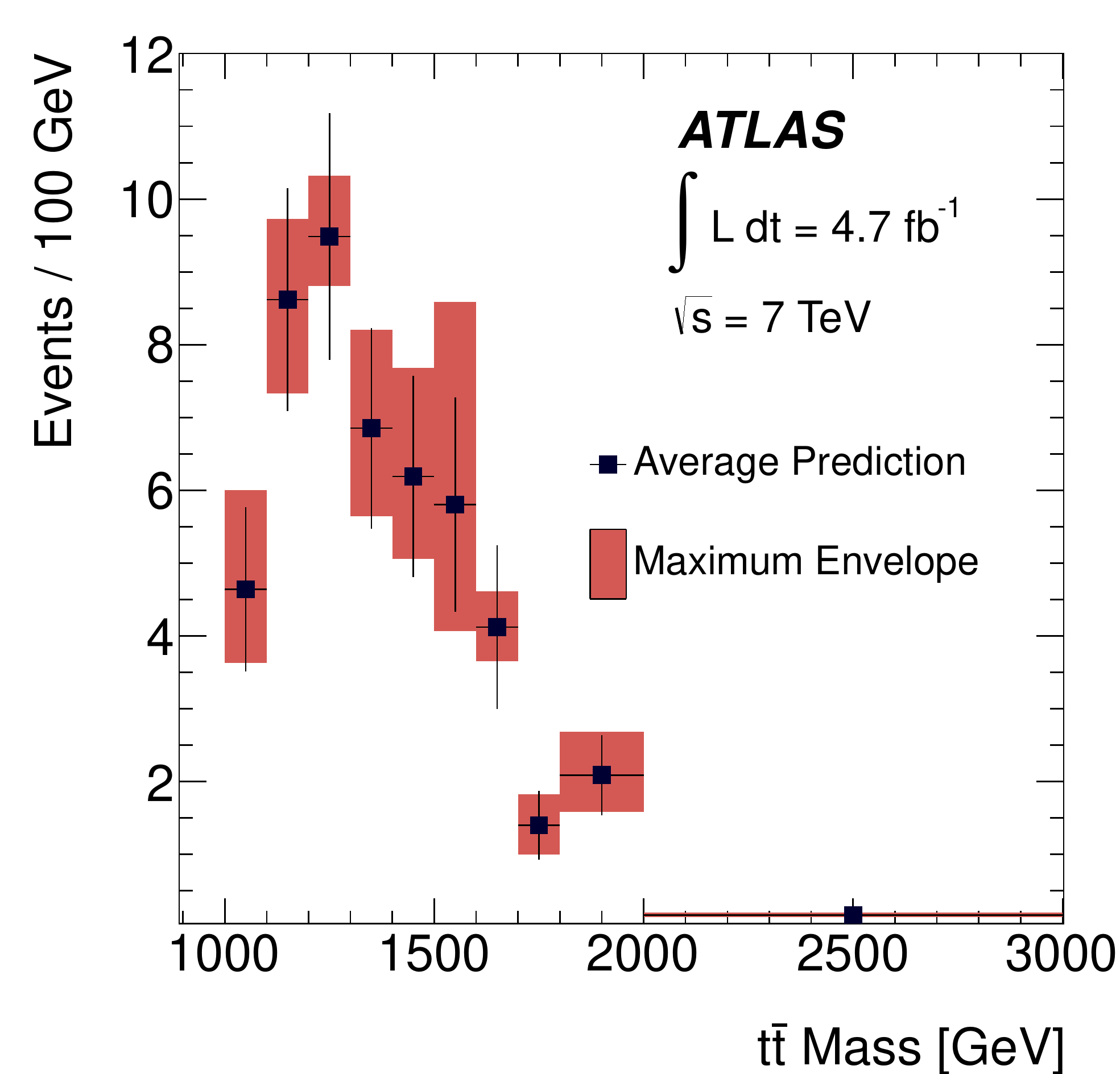}
  \caption{The data-driven prediction of the \ttbar\ mass distribution for the multijet background 
in the \ttt signal region.  
The points are the average prediction and statistical uncertainties from the five 
calculations, and the envelope is the 
range of the predictions in each bin.
}
  \label{fig:ShapeP}

\end{figure}

The prediction is verified through similar calculations using different combinations
of subsamples, as shown in Table~\ref{tab:ttt-signal-predictions}. 
The corresponding average of the predictions for the dijet mass distribution from these calculations
is shown in Fig.~\ref{fig:ShapeP}.
The results from the different calculations are in good agreement with one another, as
shown by the envelope of predictions in Fig.~\ref{fig:ShapeP}.
The averages of the individual predictions as a function of the dijet mass are used
as the estimate of the rate and shape of the multijet background in the signal region.
An independent check of this multijet background estimate is made by using the
observed rate of jets in the events prior to making the \ttt requirements,
as shown in Fig.~\ref{fig:tttjetmass}, and then using the measured rejection
of light quark/gluon jets to estimate the final background rate.
The result, $55\pm 5\ \text{(stat.)}$\ events, is in excellent agreement with the 
background estimate from the iterative calculation.

The SM \ttbar\ background in the signal region has been 
modelled using the SM Monte Carlo calculation.  
This leads to an expected yield of 
$59 ^{+27 } _{ -26 } \text{(stat.$\oplus$syst.)}$\ events. 

\begin{figure}
  \centering
  \subfigure[]{\includegraphics[width=\myFigwidth]{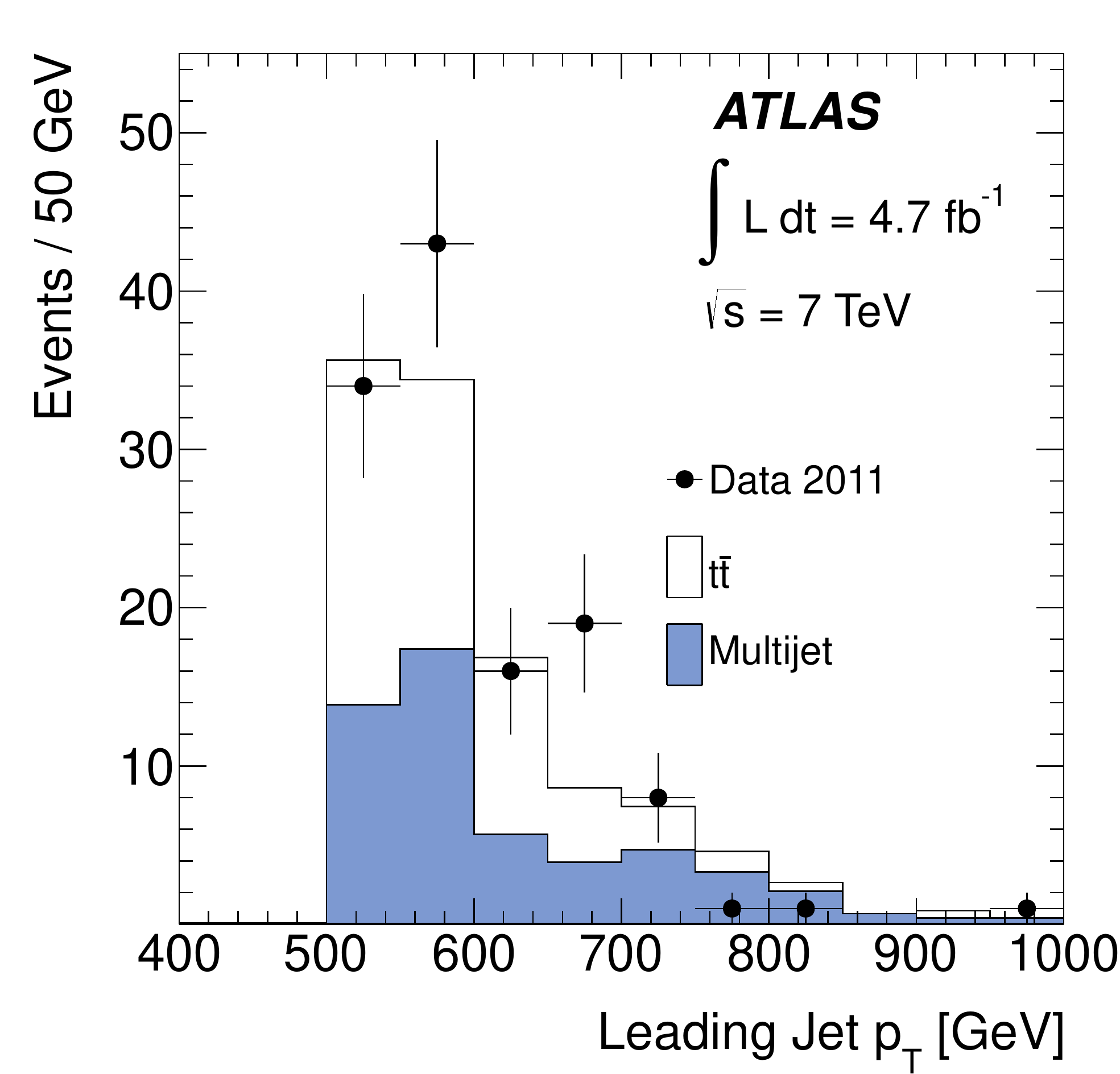}}
  \subfigure[]{\includegraphics[width=\myFigwidth]{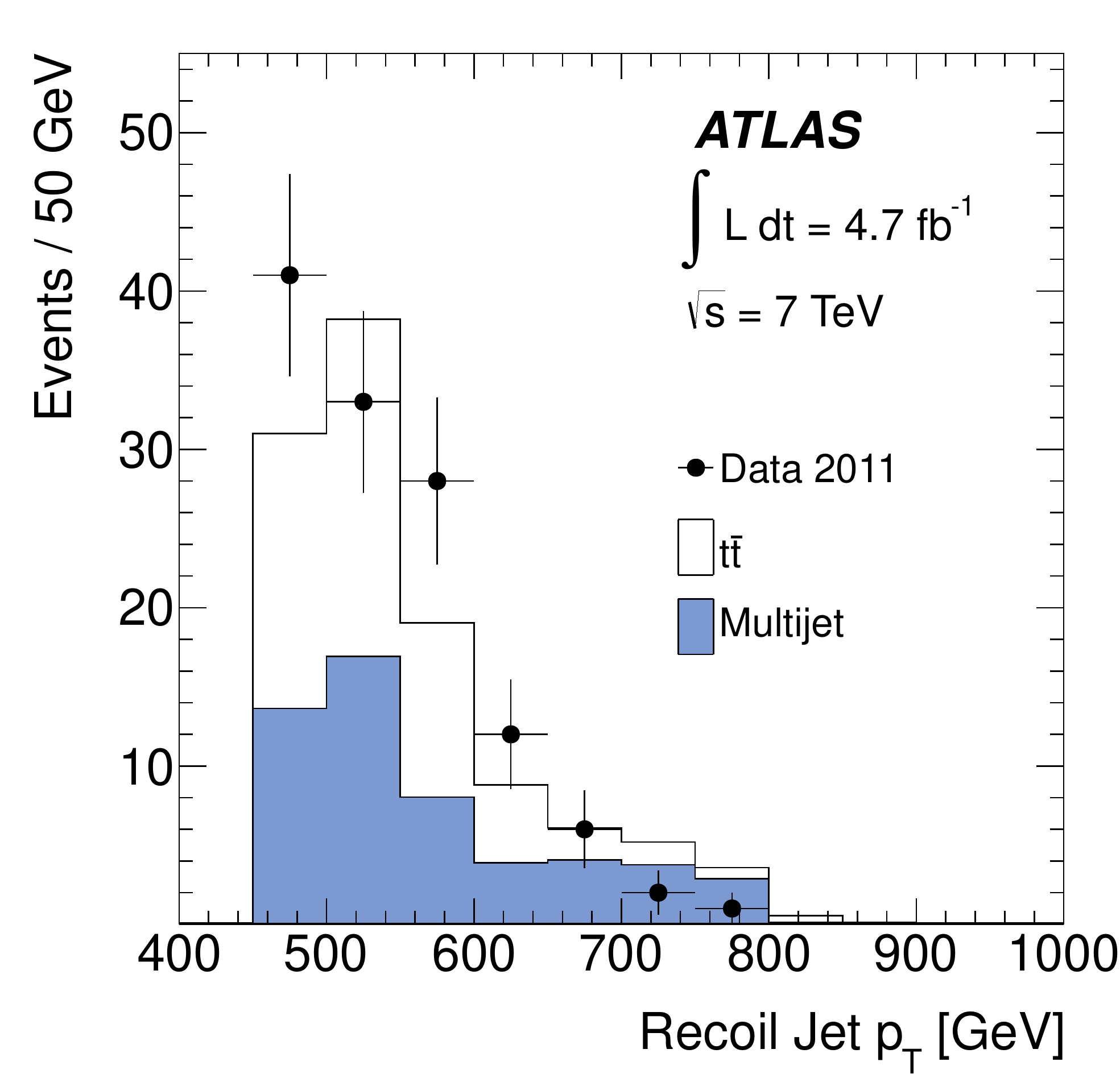}}
  \caption{Transverse momentum distributions for the leading (a) and recoil (b) jets in the \ttt signal region. 
Shown are the data distribution, the predicted SM \ttbar\ contribution and the multijet background contributions as estimated from data.
  }
  \label{fig:ttt-regionP-pt}
\end{figure}

\begin{figure}
  \centering
  \subfigure[]{\includegraphics[width=\myFigwidth]{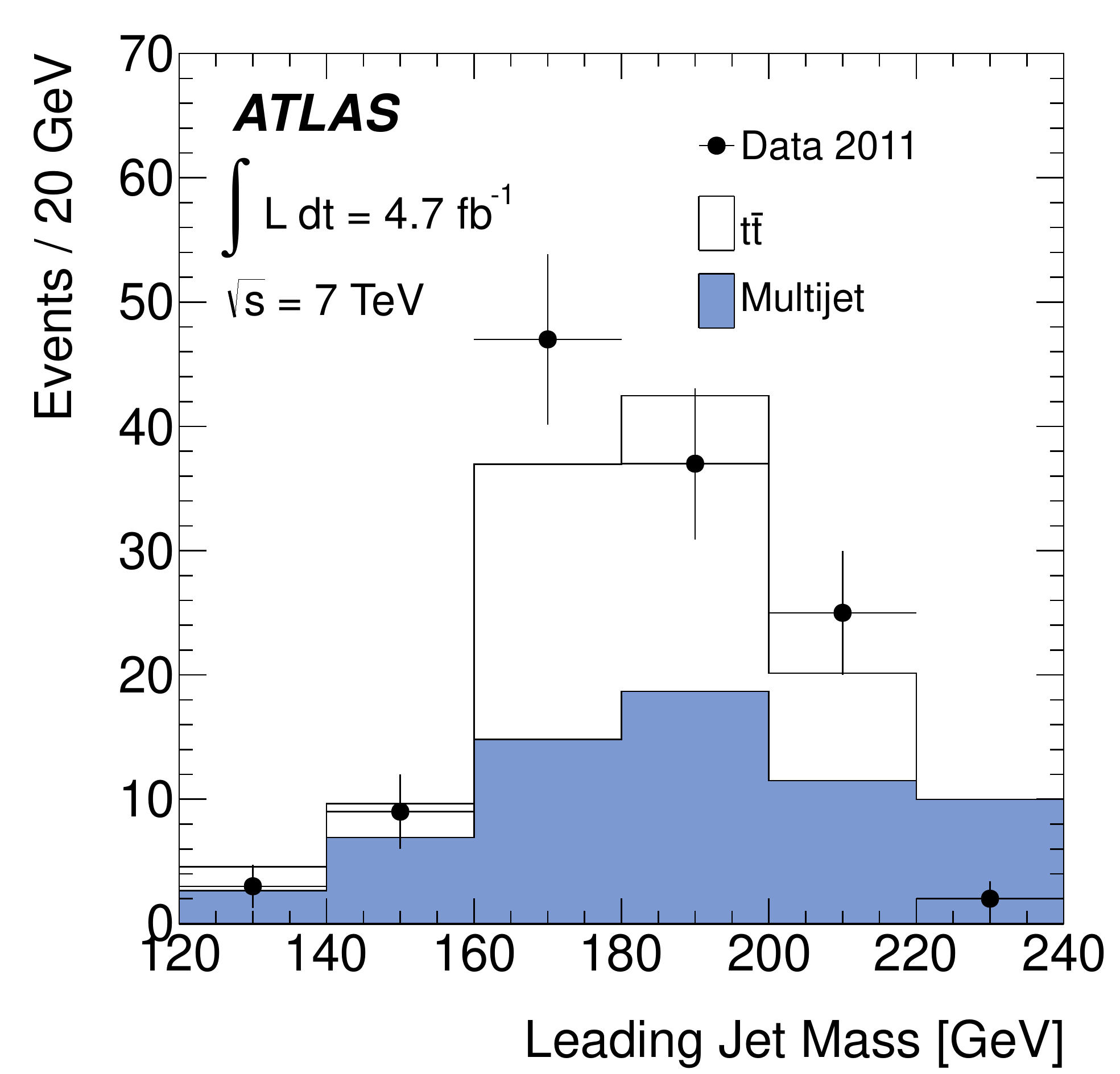}}
  \subfigure[]{\includegraphics[width=\myFigwidth]{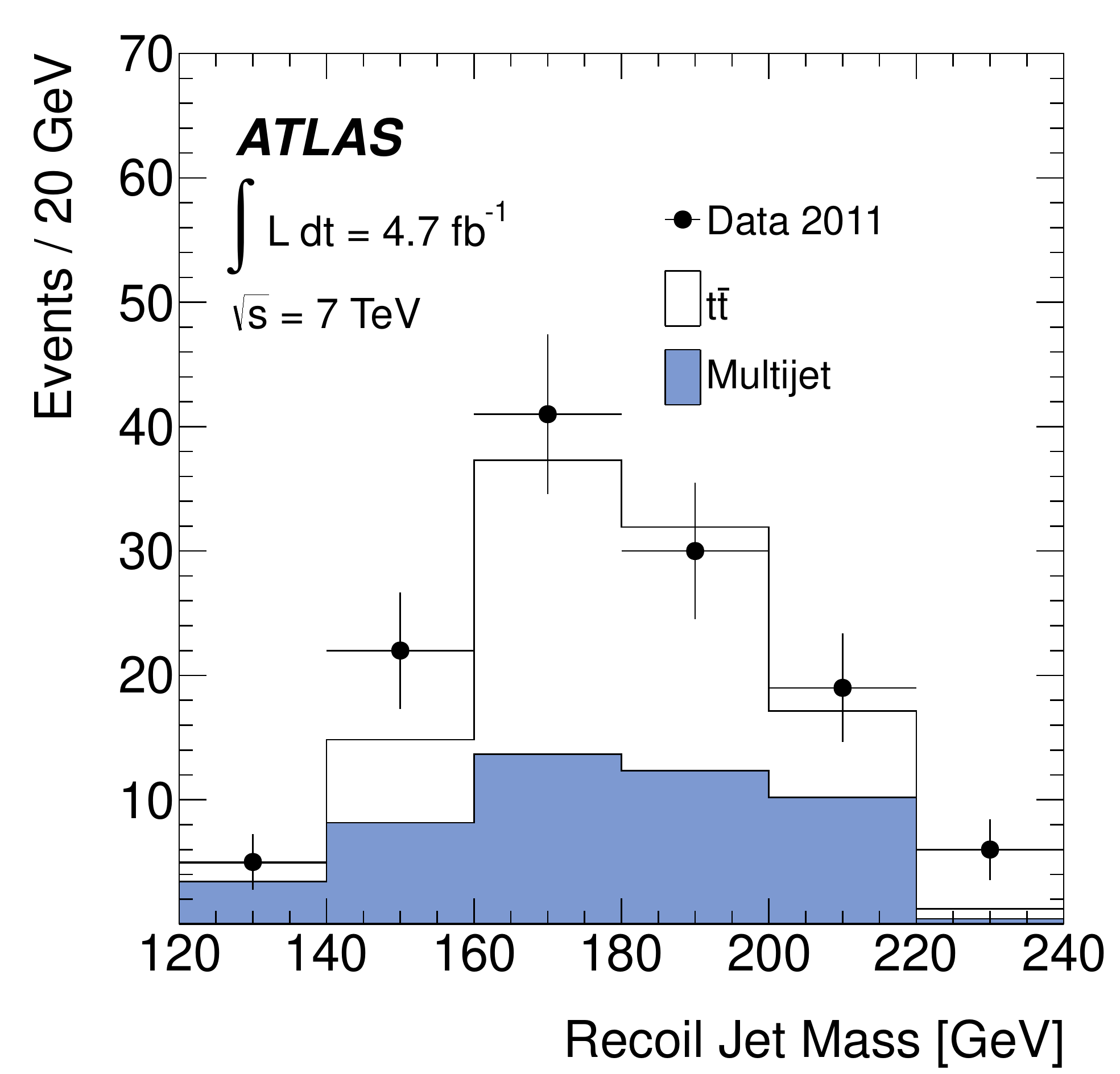}}
  \caption{
    Jet mass distributions (a) for the leading and (b) recoil jets in the \ttt signal region. 
Shown are the data distribution, the predicted SM \ttbar\ contribution and the multijet background contributions as estimated from data.
  }
  \label{fig:ttt-regionP-mass}
\end{figure}

Figs.~\ref{fig:ttt-regionP-pt}\ and \ref{fig:ttt-regionP-mass}\ show the predicted and observed \pt\ and jet mass
distributions in the \ttt signal region.
There is good agreement between the observed data and predicted background-only distributions.

\section{Systematic uncertainties}
\label{uncertainties}

The following systematic uncertainties are considered and propagated
to the predicted \mtt distributions for both analyses.
These are presented in order of their relative size, with 
the $b$-tagging efficiency and the jet energy
scale being the two largest sources of systematic uncertainty.

The uncertainty due to the $b$-tagging efficiency~\cite{AdvBtagMV1,ATLAS-CONF-2011-040,ATLAS-CONF-2012-097}\
is evaluated by re-weighting MC events according to uncertainties
on the tagging efficiency and mistag rate for
$b$-jets, $c$-jets, and light-quark and gluon jets. The $b$-tagging
efficiency has a maximum at $\pT \sim 100\GeV$. The $b$-tagging efficiency uncertainty in the
region $\pT < 200\GeV$ is determined from data 
using muon-tagged $b$-jet candidates~\cite{AdvBtagMV1}.
An additional systematic uncertainty that can be as large as $50\%$ for jets with $\pt>800\GeV$
results from limitations in the understanding of the tracking response in dense tracking environments. This additional uncertainty is added in
  quadrature to the uncertainty measured from data for lower \pT.

For the \htt\ analysis
differences in the jet energy scale (JES)
between data and simulation are determined from a comparison of the jet energy measured 
with the calorimeter and the energy measured with charged tracks associated with the jet. 
The differences vary between $2.3\%$ and $6.8\%$,
depending on the jet distance parameter, jet \pt\ and $\eta$. 
The differences have been studied independently in a sample of QCD dijet events
in which jets originate mainly from light quarks and gluons, and in a 
sample enriched in \ttbar\ events. 
For the latter sample, a lepton+jet \ttbar\
selection is made as described in Ref.~\cite{HTT-perfNote} and a fat jet
is required with $\pt > 200 \GeV$.
According to simulation this sample consists of $40\%$ \ttbar\ events.
The remaining events are characterised by the production of $W$\ bosons in association 
with light-quark and gluon jets.
This sample has a mix of quark flavours similar to the final sample in the present analysis
and also exhibits the same boosted top-quark decay topology in which the
jets are close-by. 
A similar uncertainty is found for the QCD dijet and \ttbar-enhanced samples; the maximum value is used.
The jet energy resolution (JER) for the \htt\ jets has been measured using the \pt\ asymmetry in dijet events.
The impact of differences between data and simulation is evaluated by
worsening the resolution in simulation such that it corresponds to that measured in data.

The JES uncertainty for the jets used in the \ttt analysis ranges between 4\%\ and 5\%,
depending on the jet \pt\ and $\eta$. 
The JER uncertainty has been increased by 50\%\ of that predicted by MC simulations
to account for differences in the JER measured in
data and the simulations.

The PDF eigenvector approach is applied to determine the sensitivity
of the resulting invariant mass distribution to the PDF uncertainties.
The envelope of the \CTTen~\cite{ct10}, \MSTW~\cite{mstwpdfuncert} 
and \NNPDF~\cite{nnpdf} next-to-leading-order (NLO) PDF sets is used
in this procedure~\cite{pdf4lhc}.  The uncertainty on the integrated
luminosity is $3.9\%$~\cite{lumi,ATL-2011-002}, which affects the uncertainy on the resonance 
yield and the SM \ttbar\ background.

The uncertainty due to higher-order QCD corrections to the SM
\ttbar\ background prediction is assessed by using
two alternative samples produced with the \MCatNLO\ generator in which the renormalisation and
factorisation scales have been simultaneously increased or decreased by a
factor of two.  

The impact on the shape of the $m_{\ttbar}$\ distribution of the choice of models for QCD initial and final state
radiation (ISR/FSR) and for parton showers is evaluated for the \ttbar\ sample
by comparing two different simulated samples.
The differences between the distributions are symmetrised and taken as
the systematic uncertainty. 
The variations considered are:
\begin{itemize}
  \item ISR/FSR: \ACERMC\ simulated~\cite{Kersevan:2004yg,jetveto}
    samples with two different \PYTHIA\ tunes for the simulation of
    ISR/FSR.
  \item Parton shower model: two \POWHEG\ MC \cite{FRI-0701}\ simulated samples, 
  one created using the \Herwig\ parton shower and hadronisation models  and the
other created with the \PYTHIA\ model.
\end{itemize}

The uncertainty on the $m_{\ttbar}$ distribution 
due to electroweak virtual corrections  is estimated by
adding an additional uncertainty on the SM \ttbar\ differential cross section
that is the size of the expected
reduction in the SM \ttbar\ production cross section as a function
of $m_{\ttbar}$~\cite{ewsudakov}. 

The SM \ttbar\ normalisation uncertainty is treated 
differently in the two analyses due to the different kinematic reach.
In the statistical analysis of the \htt results, 
the normalisation of the \ttbar\ contribution 
is left to be constrained in the limit-setting procedure within a variation
from $+100\%$ to $-50\%$. The width of the posterior variation is much smaller.
In the \ttt analysis the uncertainty on the SM \ttbar\ rate and 
the $m_{\ttbar}$ shape uncertainty are estimated for each systematic source.
The theoretical uncertainty on the SM \ttbar\ contribution is constrained to the
10\% uncertainty on the production cross section, convolved with the uncertainty
arising from the virtual electroweak  corrections.

The cross-checks described in Sec.~\ref{sec:HEPTopTagger}\ and Sec.~\ref{sec:TopTemplateTagger}\ show that 
the internal variables
used for the top-quark tagging methods model the data well.
In addition, as all uncertainties on the input
objects (such as the JES) are fully propagated into the two analyses
no additional uncertainty for the modelling of
top-tagging variables is added.

The trigger efficiency in the simulation is found to agree well with data, within 
the uncertainty of the jet energy scale, such that no additional trigger efficiency uncertainty
is needed.

The multijet background is estimated in a data-driven procedure that includes
subtraction of the predicted SM \ttbar\ contribution as described in Section~\ref{sec:bg}. The systematic
uncertainties on the \ttbar{} contribution are propagated to the multijet estimate.
An additional uncertainty on the multijet
background is obtained by comparing the 
$m_{\ttbar}$\ predictions using the various subsamples as described in Section~\ref{sec:bg}.

\section{Results}
\label{results}
There are 953 and 123 events observed in the \htt and \ttt signal regions, respectively. 
For the \htt selection, the SM \ttbar\
background is $770 ^{+ 220 } _{ -180 }~\text{(stat.$\oplus$syst.)}$\ events and the multijet
background is $130 \pm 70~\text{(stat.$\oplus$syst.)}$ events. 
For the \ttt selection, the SM \ttbar\
background is $59 ^{+27 } _{ -26 } \text{(stat.$\oplus$syst.)}$\ events and the multijet
background is $53 \pm 6\ \text{(stat.$\oplus$syst.)}$ events. 
The predicted SM event rates are in good agreement with the observation.

\begin{figure}
  \centering
  \subfigure[]{\includegraphics[width=\myFigwidth]{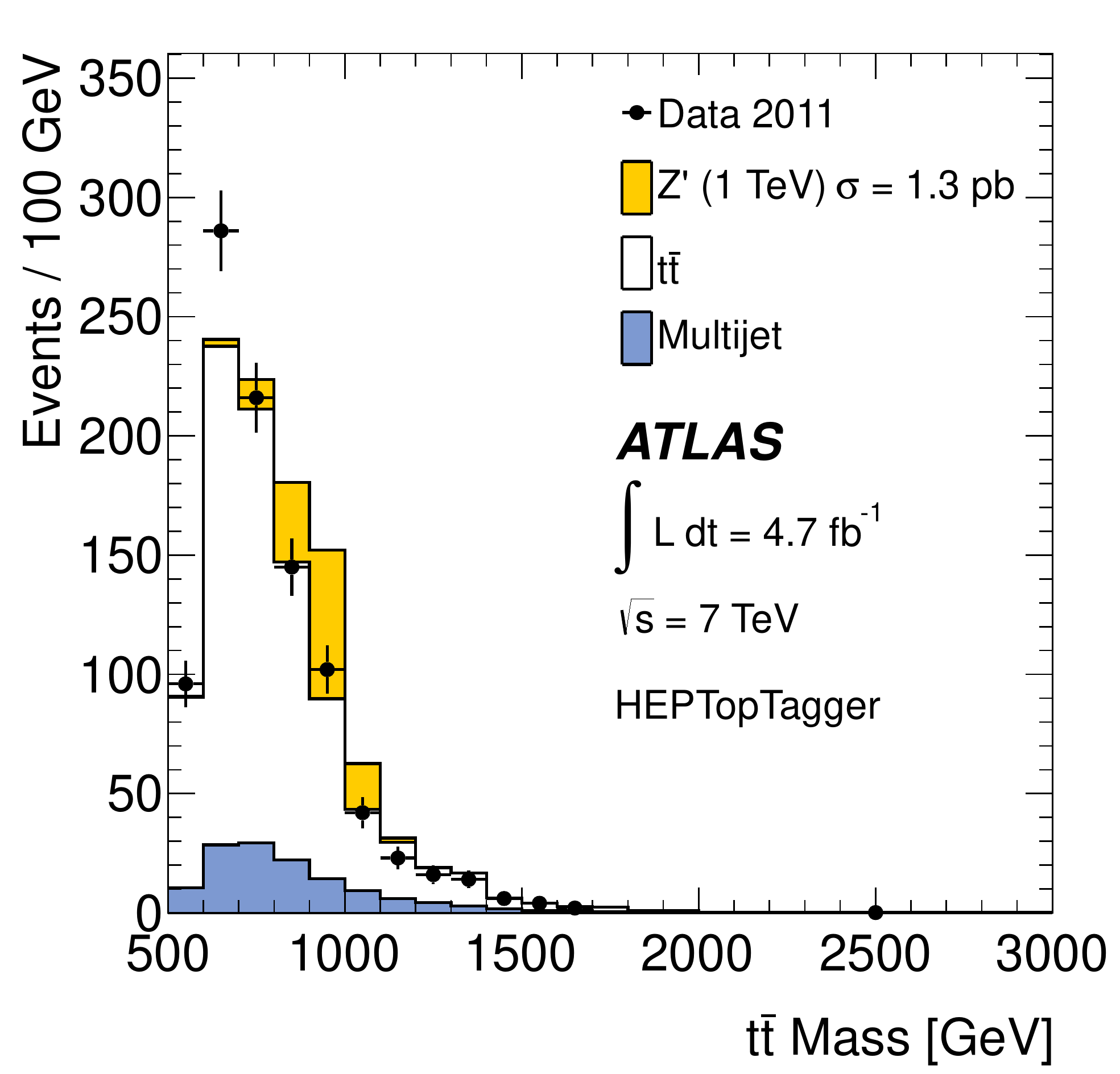}}\\
  \subfigure[]{\includegraphics[width=\myFigwidth]{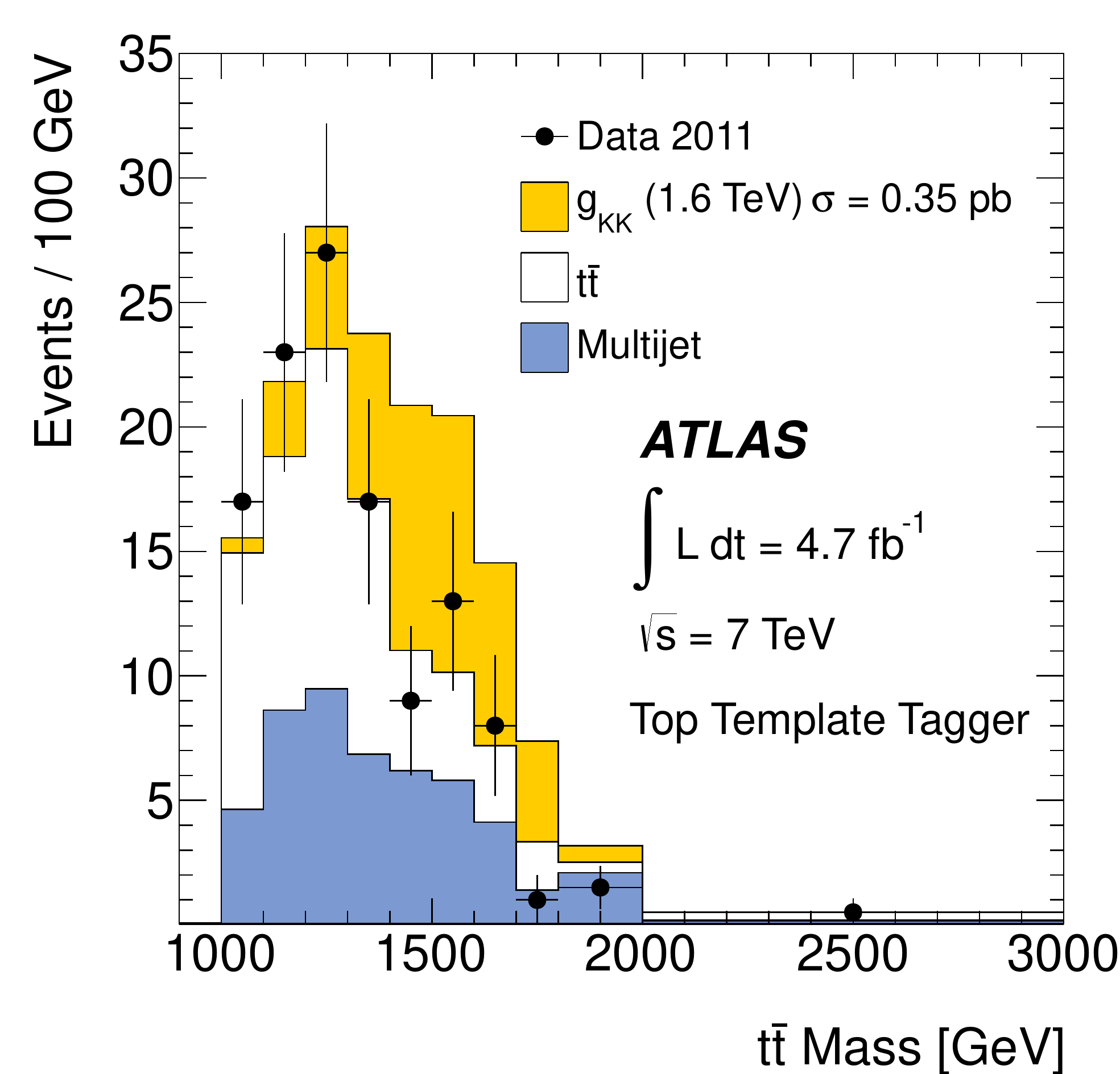}}
  \caption{Distributions of the \ttbar\ invariant mass $m_{\ttbar}$.
    The \htt data, the SM \ttbar\ background prediction, the multijet background prediction and a hypothetical \Zprime\ signal 
with $m_{\Zprime} = 1~\TeV$ are shown in (a).
The \ttt data, the SM \ttbar\ background prediction, the multijet background prediction and 
a hypothetical KK~gluon signal with
$m_{KKg}=1.6~\TeV$\ are shown in (b).
Data points show statistical uncertainties only.
}
  \label{fig:comparison-expected-mttbar}
\end{figure}

The \ttbar\ mass distributions for the data and the expected backgrounds are shown in 
Fig.~\ref{fig:comparison-expected-mttbar}.
The \ttbar\ mass binning at the lower masses is chosen to correspond approximately to the \ttbar\ mass
resolution.
For illustration, a hypothetical \Zprime\ boson signal with mass 1~\TeV\ is shown for the \htt\ \ttbar\ 
mass distribution and a hypothetical KK~gluon signal with mass 1.6~\TeV\ is shown for the
\ttt\ \ttbar\ mass distribution.
No statistically significant excess over the SM \ttbar\ expectation plus multijet background is observed at any mass value.

As no signal is observed in either selection, 95\%\ 
CL upper limits are set on the production cross section times
branching ratio to \ttbar\ final states for each model using a Bayesian approach~\cite{Bertram:2000br}.
A binned likelihood function based on Poisson distributions for each \ttbar\ invariant mass bin is used.

The limits are determined for resonance masses ranging from 0.5 to 2.0~\TeV\ 
for the \Zprime\ boson model 
and 0.7 to 2.0~\TeV\ for the KK gluon model.
The systematic uncertainties are treated as nuisance
parameters with Gaussian prior distributions reflecting their
uncertainty and are then marginalised to set credibility intervals.

The large uncertainty on the SM \ttbar\ normalisation in the \htt selection
by construction precludes other nuisance
parameters that are sensitive to this normalisation
to be strongly constrained.
To prevent regions with low $m_{\ttbar}$, where  high
event yields result in small statistical uncertainties, constraining regions with high
$m_{\ttbar}$, 
the jet energy scale uncertainty is treated as being uncorrelated
 between different bins in jet~\pT.
Studies of the posterior distributions of the nuisance parameters have been
performed to ensure that the uncertainties arising from the parton
shower model and ISR/FSR do not over-constrain the uncertainties.

To estimate the {\it a priori}\ sensitivity of this search, background-only 
pseudo-experiments are randomly drawn from the background prediction.  All 
nuisance parameters are allowed to vary in a manner consistent with their prior distributions
for each pseudo-experiment. 

The median of the distribution is chosen to represent the expected limit. 
The ensemble of limits
is also used to define the 68\%\ and 95\%\ CL envelope of limits as a function of resonance mass.  

The dominant systematic uncertainties in both analyses come from the uncertainties on
$b$-tagging efficiency, jet energy scale and SM \ttbar\ normalisation.

\begin{figure}[ht]
  \centering
  \subfigure[]{\includegraphics[width=\myFigwidth]{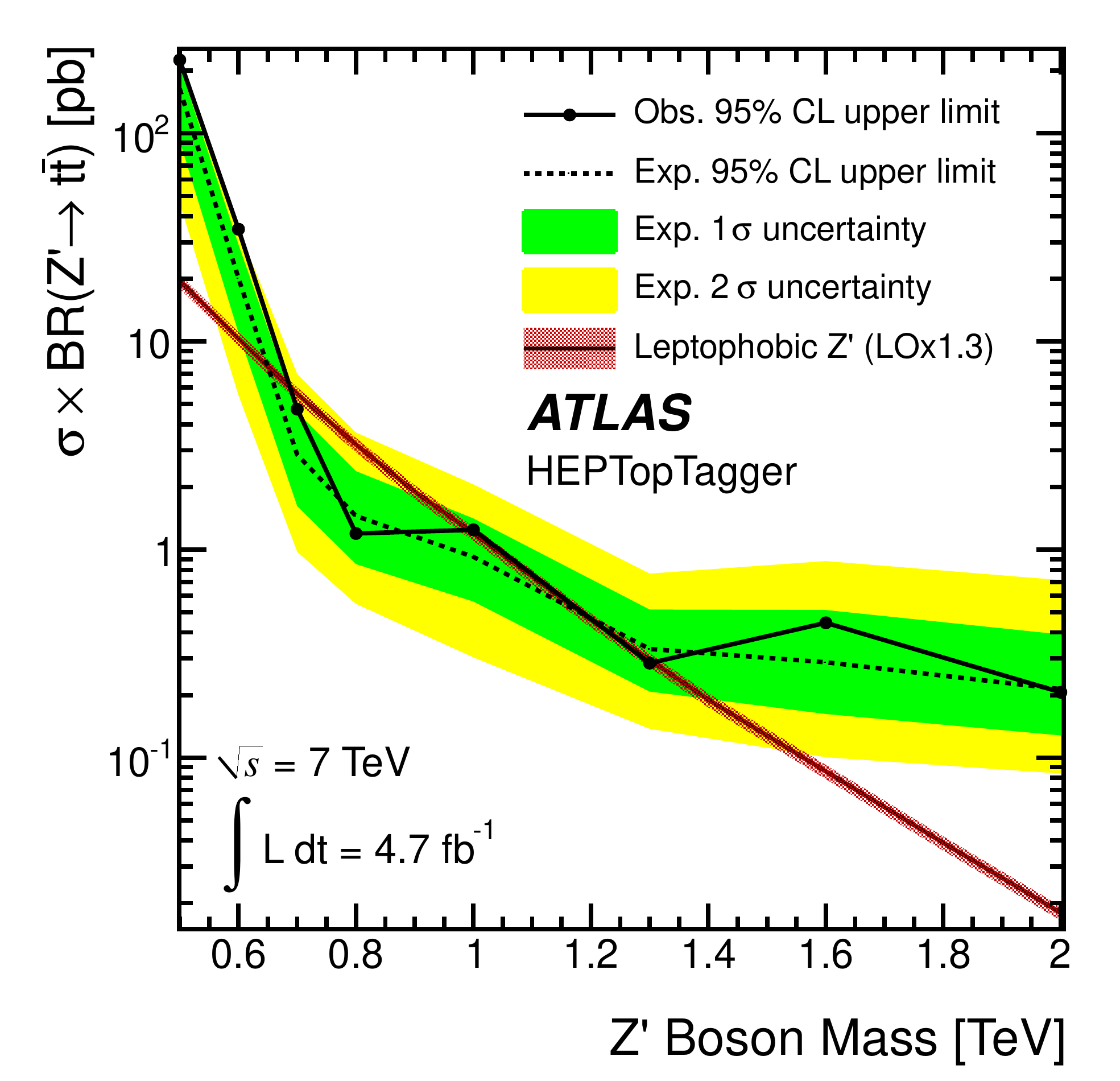}}\\
  \subfigure[]{\includegraphics[width=\myFigwidth]{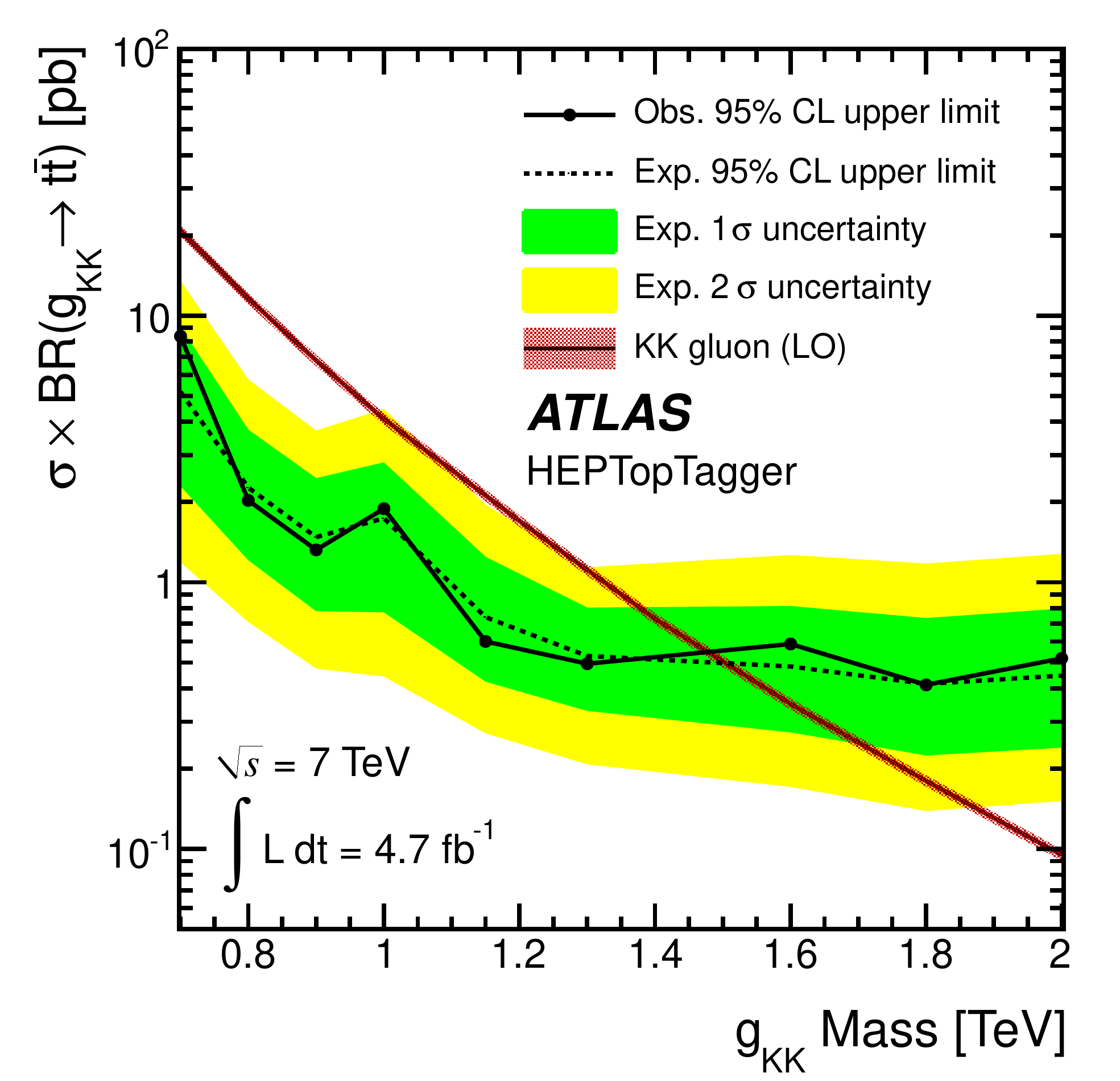}}

  \caption{Expected and observed 95\%\ CL upper limits on the production cross section times branching fraction \xbr\ 
 as a function of (a) the \Zprime\ boson mass and
(b)    the KK~gluon mass for the \htt\ selection. 
The red bands are the model
    predictions including theoretical uncertainties. 
The \Zprime\ boson leading-order (LO) cross section is 
    multiplied by 1.3 to account for expected higher-order
    corrections. The KK~gluon LO cross section is used.}
  \label{fig:htt-limits}
\end{figure}

\begin{figure}[h!]
  \centering
  \subfigure[]{\includegraphics[width=\myFigwidth]{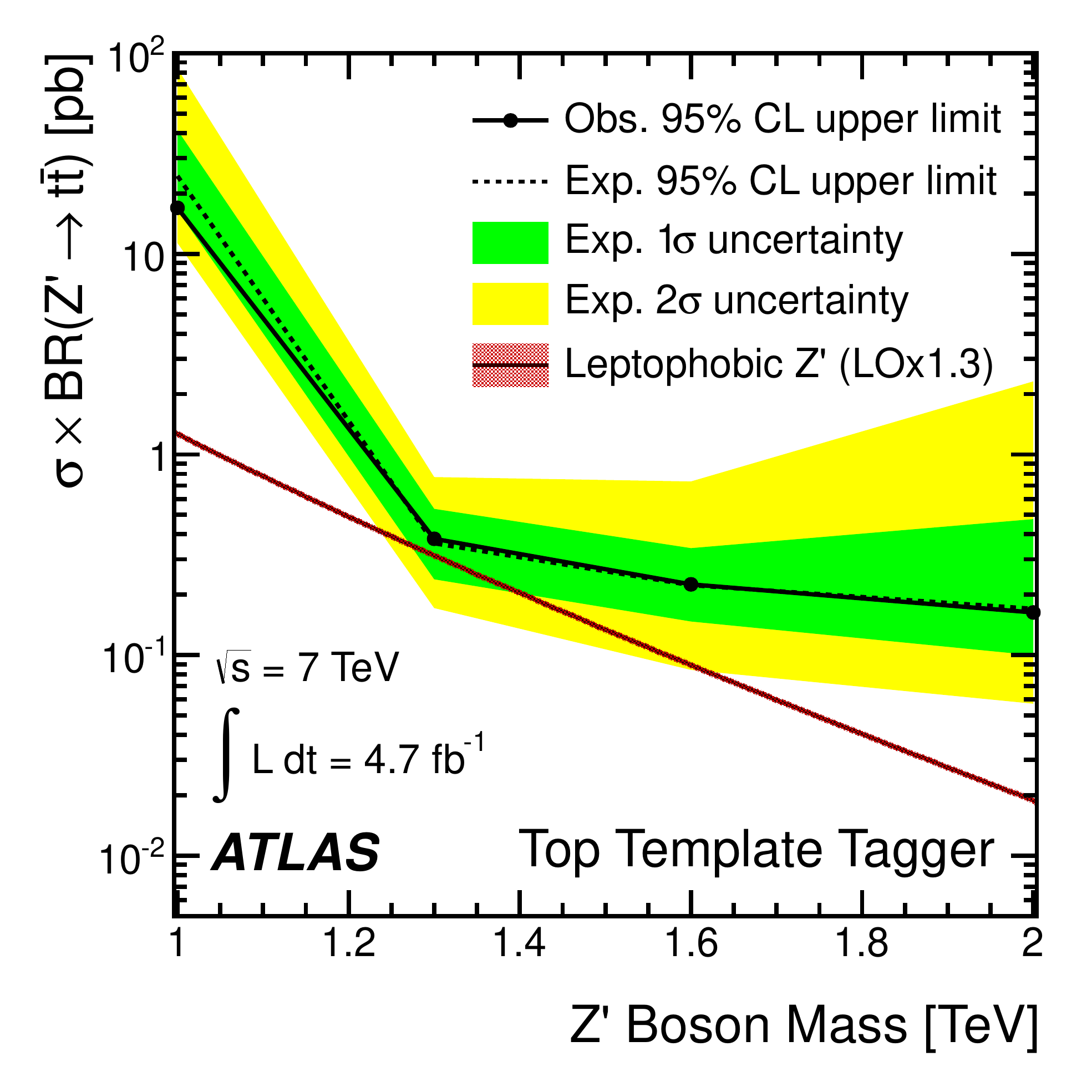}}
  \subfigure[]{\includegraphics[width=\myFigwidth]{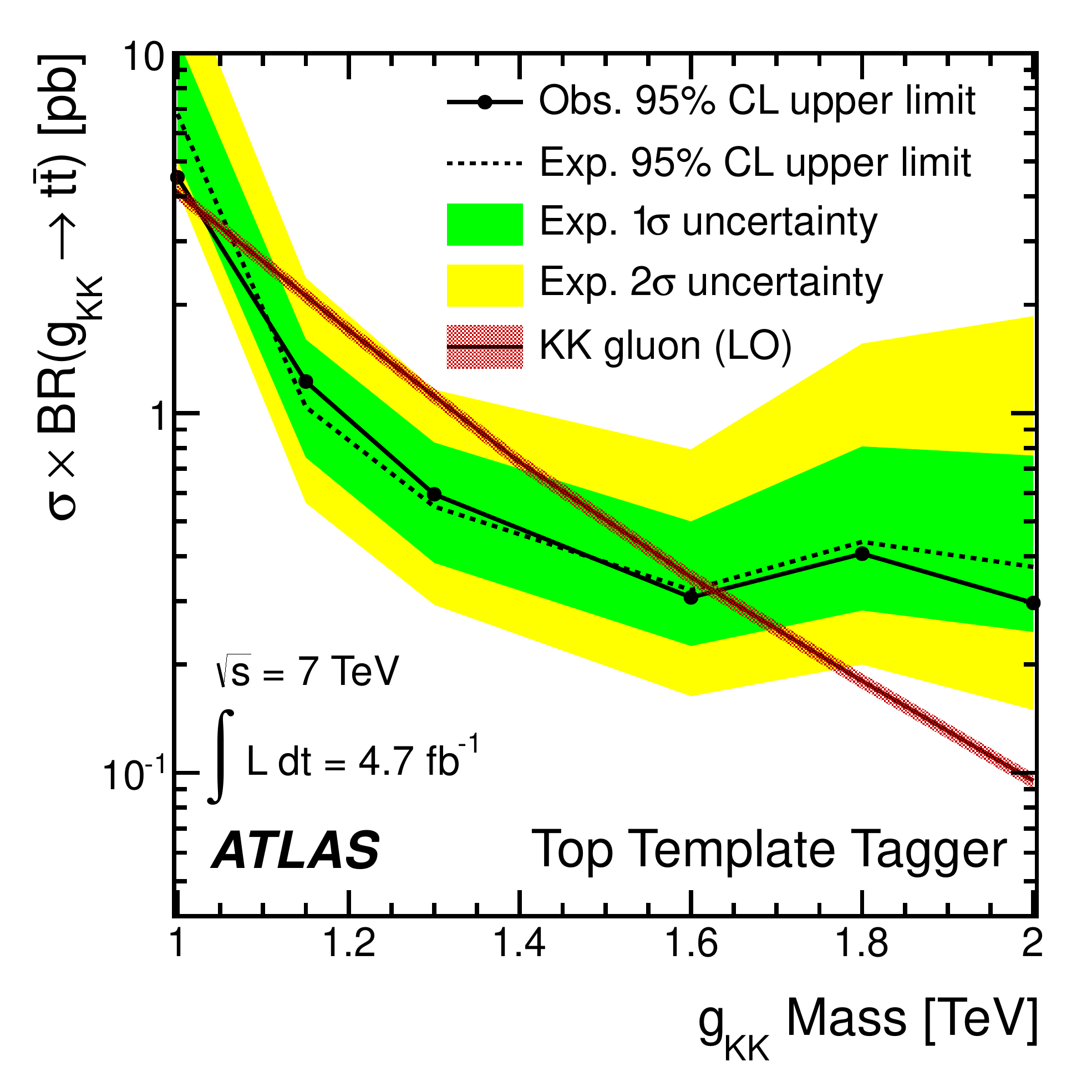}}\\

  \caption{
  Expected and observed 95\%\ CL upper limits on the production cross section times branching fraction \xbr\
 as a function of (a) the \Zprime\ boson mass and (b) the KK~gluon mass for the \ttt selection.
The red bands are the model predictions including theoretical uncertainties.
The \Zprime\ boson LO cross section is 
multiplied by 1.3 to account for expected higher order
corrections. The KK~gluon LO cross section is used.}
  \label{fig:ttt-limits}
\end{figure}

Figs.~\ref{fig:htt-limits} and~\ref{fig:ttt-limits}\ show the \htt and \ttt 95\% CL exclusion
limits on the cross section times branching ratio for the two models. 
They are interpreted as mass limits by comparing  the cross-section limits to theoretical cross-section 
predictions as a function of mass from specific benchmark models.  
The expected and observed mass limits are shown in
Table~\ref{tab:massLimits}. 

As described in Ref.~\cite{colorstruct}, the colour structure of the
KK resonance can affect the tagging efficiency. 
This effect is small, but the
results presented here are valid only for resonances with the same colour
structure as the KK~gluon 
(e.g., the sensitivity for a KK~photon with the same mass and
width as a KK~gluon will differ by $\approx 10\:\%$).

\begin{table}[!thb]
  \centering
  \caption{Expected (Exp.) and observed (Obs.) exclusion regions on
    the leptophobic \Zprime\ boson mass and on the KK gluon mass in
    the Randall--Sundrum model.}
\vskip 0.2cm
  \begin{tabular}{|l|c|c|}
   \hline
    Model  & Obs. Limit (\TeV) & Exp. Limit (\TeV)\\
    \hline
\multicolumn{3}{|c|}{\htt} \\
\hline
    \Zprime  & $0.70 < m_{\Zprime} < 1.00$ & $0.68  < m_{\Zprime} < 1.16$ \\
             & $1.28 < m_{\Zprime} < 1.32$ & \\
    KK gluon         & $0.70 < m_{\gKK} < 1.48$ & $0.70  < m_{\gKK} < 1.52$  \\
\hline
\multicolumn{3}{|c|}{\ttt} \\
\hline
    KK gluon         & $1.02 < m_{\gKK} < 1.62$ & $1.08  < m_{\gKK} < 1.62$  \\
    \hline
  \end{tabular}
 \label{tab:massLimits}
\end{table}

The data samples for the two analyses are statistically correlated.  
However, the expected limits are different for the two analyses and illustrate 
their complementarity: 
The \htt selection is able to exclude \Zprime\ boson resonances over part of the mass
range between 0.70 and 1.32~\TeV\ and KK gluons with masses between 0.70 and 1.48~\TeV.
The \ttt selection is not able to set an exclusion limit on \Zprime\ boson resonances but 
is able to exclude the wider-width KK~gluon resonances for masses between 1.02 and 1.62~\TeV.

To combine the limits from these two analyses, the results from the 
tagger with the lower expected exclusion limit are selected.
The \htt selection provides lower expected limits for \Zprime\ boson masses 
up to 1.3~\TeV, and for KK~gluons with masses between 0.7 and 1.3~\TeV.
The \ttt selection provides the lower expected limits for both \Zprime\ bosons and KK~gluons 
with masses above 1.4~\TeV.
These two analyses together are able to exclude the \Zprime\ boson model with masses
$0.70<m_{\Zprime} < 1.00$~\TeV\ and $1.28<m_{\Zprime}<1.32$~\TeV, and KK~gluons with masses 
$0.70 < m_{g_{KK}} < 1.62$~\TeV, all at 95\%\ CL.

\section{Conclusions}

A search for massive resonances, characterised by a narrow state such as a 
\Zprime\ boson or a wider object such as a KK~ gluon, 
decaying into \ttbar\ pairs in the fully hadronic final state is presented. 
The analysis uses a dataset corresponding to $4.7\:\text{fb}^{-1}$, collected with the ATLAS
detector during the 2011 $pp$ run of the LHC at a
centre-of-mass energy of $7\:\text{TeV}$. 
Two top-quark tagging schemes, the \htt and \ttt methods, are used to identify and reconstruct 
top-quark pairs in their hadronic decay mode for boosted top quarks 
with transverse momenta between 200~\GeV\ and approximately 1~\TeV.

The reconstructed $m_{\ttbar}$ spectra are compared to predictions for SM \ttbar\ production and
background from massive jets 
produced through QCD interactions. 
No evidence for resonant \ttbar\ production is found using either top-quark tagging method. 
These two analyses together exclude the \Zprime\ boson model with masses
$0.70<m_{\Zprime} < 1.00$~\TeV\ and $1.28<m_{\Zprime}<1.32$~\TeV, and KK~gluons with masses 
$0.70 < m_{g_{KK}} < 1.62$~\TeV, all at 95\%\ CL.
These results extend the previous ATLAS limits on \Zprime\ bosons and KK~gluons production that
were based on the lepton plus jets final state.


\section{Acknowledgements}

We acknowledge the contributions of
Jose Juknevich and Mihailo Backovic, Weizmann Institute of Science,
for insights and calculations.

We thank CERN for the very successful operation of the LHC, as well as the
support staff from our institutions without whom ATLAS could not be
operated efficiently.

We acknowledge the support of ANPCyT, Argentina; YerPhI, Armenia; ARC,
Australia; BMWF and FWF, Austria; ANAS, Azerbaijan; SSTC, Belarus; CNPq and FAPESP,
Brazil; NSERC, NRC and CFI, Canada; CERN; CONICYT, Chile; CAS, MOST and NSFC,
China; COLCIENCIAS, Colombia; MSMT CR, MPO CR and VSC CR, Czech Republic;
DNRF, DNSRC and Lundbeck Foundation, Denmark; EPLANET, ERC and NSRF, European Union;
IN2P3-CNRS, CEA-DSM/IRFU, France; GNSF, Georgia; BMBF, DFG, HGF, MPG and AvH
Foundation, Germany; GSRT and NSRF, Greece; ISF, MINERVA, GIF, DIP and Benoziyo Center,
Israel; INFN, Italy; MEXT and JSPS, Japan; CNRST, Morocco; FOM and NWO,
Netherlands; BRF and RCN, Norway; MNiSW, Poland; GRICES and FCT, Portugal; MERYS
(MECTS), Romania; MES of Russia and ROSATOM, Russian Federation; JINR; MSTD,
Serbia; MSSR, Slovakia; ARRS and MVZT, Slovenia; DST/NRF, South Africa;
MICINN, Spain; SRC and Wallenberg Foundation, Sweden; SER, SNSF and Cantons of
Bern and Geneva, Switzerland; NSC, Taiwan; TAEK, Turkey; STFC, the Royal
Society and Leverhulme Trust, United Kingdom; DOE and NSF, United States of
America.

The crucial computing support from all WLCG partners is acknowledged
gratefully, in particular from CERN and the ATLAS Tier-1 facilities at
TRIUMF (Canada), NDGF (Denmark, Norway, Sweden), CC-IN2P3 (France),
KIT/GridKA (Germany), INFN-CNAF (Italy), NL-T1 (Netherlands), PIC (Spain),
ASGC (Taiwan), RAL (UK) and BNL (USA) and in the Tier-2 facilities
worldwide.

\clearpage

\bibliographystyle{JHEP.bst} 

\providecommand{\href}[2]{#2}\begingroup\raggedright\endgroup

\clearpage 

\begin{flushleft}
{\Large The ATLAS Collaboration}

\bigskip

G.~Aad$^{\rm 48}$,
T.~Abajyan$^{\rm 21}$,
B.~Abbott$^{\rm 111}$,
J.~Abdallah$^{\rm 12}$,
S.~Abdel~Khalek$^{\rm 115}$,
A.A.~Abdelalim$^{\rm 49}$,
O.~Abdinov$^{\rm 11}$,
R.~Aben$^{\rm 105}$,
B.~Abi$^{\rm 112}$,
M.~Abolins$^{\rm 88}$
% ...
\bigskip
\\
% ...
$^{11}$ Institute of Physics, Azerbaijan Academy of Sciences, Baku, Azerbaijan\\
$^{12}$ Institut de F{\'\i}sica d'Altes Energies and Departament de F{\'\i}sica de la Universitat Aut{\`o}noma de Barcelona and ICREA, Barcelona, Spain\\
% ...
$^{21}$ Physikalisches Institut, University of Bonn, Bonn, Germany\\
% ...
$^{48}$ Fakult{\"a}t f{\"u}r Mathematik und Physik, Albert-Ludwigs-Universit{\"a}t, Freiburg, Germany\\
$^{49}$ Section de Physique, Universit{\'e} de Gen{\`e}ve, Geneva, Switzerland\\
% ...
$^{88}$ Department of Physics and Astronomy, Michigan State University, East Lansing MI, United States of America\\
% ...
$^{105}$ Nikhef National Institute for Subatomic Physics and University of Amsterdam, Amsterdam, Netherlands\\
% ...
$^{111}$ Homer L. Dodge Department of Physics and Astronomy, University of Oklahoma, Norman OK, United States of America\\
$^{112}$ Department of Physics, Oklahoma State University, Stillwater OK, United States of America\\
% ...
$^{115}$ LAL, Universit{\'e} Paris-Sud and CNRS/IN2P3, Orsay, France\\
% ...
\end{flushleft}


\end{document}